\documentstyle[Sprocl,epsf,epsfig]{article}
\def\be{\begin{equation}}
\def\ee{\end{equation}}
\def\bea{\begin{eqnarray}}
\def\eea{\end{eqnarray}}

\def\GeV{\hbox{$\;\hbox{\rm GeV}$}}
\def\TeV{\hbox{$\;\hbox{\rm TeV}$}}

\newcommand{\picob}{\mbox{{\rm ~pb}}}
\newcommand{\Rp}{\mbox{$\not \hspace{-0.15cm} R_p$}}

\newcommand{\lsim}{\raisebox{-1.5mm}{$\:\stackrel{\textstyle{<}}{\textstyle{\sim}}\:$}}
\newcommand{\gsim}{\raisebox{-0.5mm}{$\stackrel{>}{\scriptstyle{\sim}}$}}
\begin{document}

\vspace*{-1.3cm}

\noindent
{\it To be published in the proceedings of the WEIN 1998 Conference}
\vspace*{0.4cm}

\title{SEARCHES FOR NEW BOSONS COUPLING TO $e-q$ PAIRS AT HERA 
       AND OTHER COLLIDERS}
\author{ Y. SIROIS }
\address{LPNHE Ecole Polytechnique, IN2P3-CNRS, \\
         Palaiseau, France \\
         E-mail: sirois@polhp2.in2p3.fr}
\maketitle
%========================= ABSTRACT ====================================
\abstracts{The early observation at HERA of an excess of events compared  
	   to the expectation from  the Standard Model in very short 
	   distance $e^+p$ deep-inelastic scattering processes 
	   has renewed the interest in the search for new physics which 
	   could manifest in electroweak-like interactions.
           New  preliminary results from the H1 and ZEUS experiments  
	   making use of all available $e^+p$ data are reviewed here, 
	   with an emphasis on the search for new bosons possessing 
	   Yukawa couplings to lepton-quark pairs.
           The sensitivity of HERA to leptoquarks, and to squarks of 
	   R-parity violating supersymmetry, is confronted to existing 
	   indirect constraints from rare and forbidden semi-leptonic 
	   decays, atomic parity violation and neutrinoless double-beta 
	   decay, as well as to direct constraints from LEP and Tevatron
	   colliders.
           The HERA and Tevatron colliders are found to offer exciting  
	   prospects for new physics, accessing yet unexplored domains 
	   of the mass-coupling plane.
           Possible striking manifestation of explicit lepton flavour 
	   violation is also discussed.}
%
%=======================================================================
\section{Introduction}
\label{sec:intro}

The interest in the search for a physics beyond the Standard Model (SM) 
which could interfere with electroweak-like interactions has been
considerably enhanced recently by the observation in the H1~\cite{H1HIQ2} 
and ZEUS~\cite{ZEUSHIQ2} experiments at HERA of a deviation from
SM expectations in neutral current (NC) deep-inelastic scattering (DIS) 
events at very high squared momentum-transfer $Q^2$ and possibly, 
yet at a less significant level, in charged current (CC) processes. 
In particular, an apparent ``clustering'' of outstanding NC-like events at 
masses around $200 \GeV$ has motivated considerable work on leptoquark 
constraints and phenomenology~\cite{HIGHXYLQ} and on squarks in 
$\Rp$-SUSY~\cite{HIGHXYRPV}, while contact interactions were also discussed 
as a possible source of distortion of the high $Q^2$ spectra~\cite{HIGHXYCI}.

The $ep$ collider HERA offers unique possibilities to search for 
$s$-channel production of new scalar particles which couple to 
lepton-parton pairs, up to a kinematic limit of 
$\sqrt{s_{ep}} \simeq 300 \GeV$. 
The mass range relevant for a possible discovery at HERA has now been 
severely constrained by the TeVatron experiments~\cite{D01GENE,CDF1GENE}
in particular for scalars decaying with a large branching ratio 
$\beta_{eq}$ into electron$+$quark. 
These constraints can be partly avoided for the squarks in $\Rp$-SUSY theories
where $\beta_{eq}$ is naturally small given the competition with gauge 
decay modes. 
Exhaustive squark searches covering both $\Rp$ decay modes and various 
possible gauge decay modes are thus strongly motivated.

The original H1 and ZEUS results mentioned above were based on data samples 
collected from 1994 to 1996.
The analyses have now been updated incorporating also the 1997 data,
for a total gain of a factor $\simeq 2.5$ in integrated luminosity. 
New preliminary inclusive NC and CC HERA results using all available $e^+p$ 
data will be reviewed in section~\ref{sec:highq2} and~\ref{sec:xsect}
of this paper. Interpretation and constraints on contact interactions
will be summarized in section~\ref{sec:contact}.
New results on leptoquark searches will be discussed in 
section~\ref{sec:leptoquark} and searches for squarks of $\Rp$-SUSY 
in section~\ref{sec:rpvsusy}.
Searches for direct lepton flavour violating processes will be 
discussed in the context of both leptoquark phenomenology and
$\Rp$-SUSY. The HERA results and prospects will be compared to those 
of indirect processes and to direct searches at other colliders.

%=======================================================================
\section{New Physics and Very High $Q^2$ Rates at HERA}
\label{sec:highq2}

% Introduction:

The investigations of very high $Q^2$ DIS-like processes at HERA, even 
with low statistics, are strongly motivated by the potential reach for 
new physics beyond the SM such as the production of new scalar ($X_S$)
resonances. 
An $X_S$ produced in the $s$-channel could for example lead to individual 
event signatures indistinguishable from standard NC and CC DIS if it
decays into $e+q$ or $\nu+q$. 
The new signal would nevertheless be identified statistically as a peak in 
the invariant mass distribution, associated with a characteristic angular 
distribution of the decay products. 
As will be discussed in section~\ref{sec:leptoquark}, the $X_S$ particles
which decay uniformly in their CM frame would contribute most significantly 
at large $Q^2$ or large $y$ where $Q^2= M^2 y$, $M$ is the resonance
mass and $y$ is related the decay polar angle of the final state lepton.
Such new bosons possessing Yukawa couplings to lepton-quark pairs can in
addition contribute in the $u$-channel by converting a lepton into a quark
(and vice versa). Interference with SM DIS processes could also originate
from effective contact interaction caused by the exchange of new bosons
of masses $M \gg \sqrt{s_{ep}}$.

It is with these considerations in mind that H1 and ZEUS experiments have 
carried their original analysis~\cite{H1HIQ2,ZEUSHIQ2} of the 
1994$\rightarrow$96 data corresponding to integrated
luminosities of ${\cal{L}}_{H1} \simeq 14.2 \pm 0.3 \picob^{-1}$ and
${\cal{L}}_{ZEUS} \simeq 20.1 \pm .5 \picob^{-1}$. 
The detailed comparison of the high $Q^2$ tail of NC and CC-like event
rates with expectation from SM DIS processes has revealed exciting 
features.
%
% ------------- FIGURE 1: Q2 plot y > 0.1 NC selection  ----------------
%
\begin{figure}[htb]

  \vspace*{-1.0cm}

  \begin{center}
   \epsfxsize=0.70\textwidth 
    \epsffile{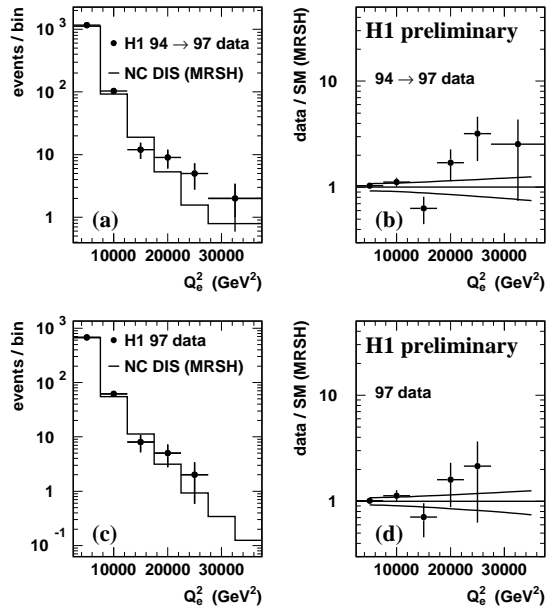}

   \caption[]{ \label{fig:q2rates}
      { \small  (a) $Q^2_e$ distribution of the NC DIS candidate 
                events for the 1994$\rightarrow$97 H1 data ($\bullet$) and 
		for standard NC DIS expectation (histogram);
                (b) ratio of the observed and expected number of
                events as a function of $Q^2_e$; 
                the lines above and below unity specify the 
                $\pm 1\sigma$ levels determined using the combination 
                of statistical and systematic errors of 
                the NC DIS expectation;
                (c) and (d) : as (a) and (b) but for 1997 data alone. }}
 \end{center}
\end{figure}
%------------------------------------------------------------------------
These observations have now been updated including 1997 data, for a
total of ${\cal{L}}_{H1} \simeq 37 \picob^{-1}$ and
${\cal{L}}_{ZEUS} \simeq 46.6 \picob^{-1}$.
I shall now briefly review the essential features of the original and
updated rate measurements.
The results obtained when ``converting'' the inclusive DIS measurements
into differential cross-sections will be discussed in the next section.

The selection of NC-like events at HERA is straightforward. 
It essentially requires an isolated $e^{\pm}$ at high $E_{T,e}$ well
balanced by the hadronic transverse flow and with no longitudinal 
losses ``visible'' along the $e$ beam $-z$ direction 
($\sum_{vis.} E-P_z \simeq 2 E_e^{beam}$).
% --------------- FIGURE : ----------------------------------------------
\begin{figure}[htb]
  \begin{center}
   \epsfxsize=0.80\textwidth 
    \epsffile{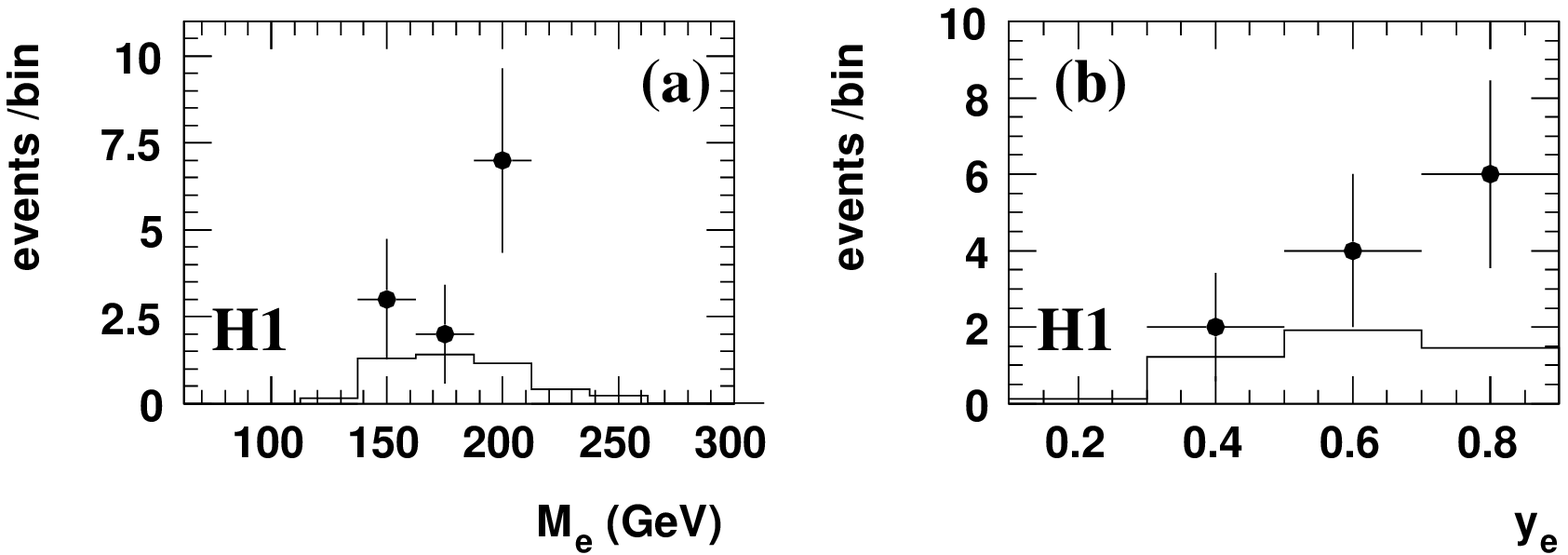}
   \epsfxsize=0.80\textwidth 
    \epsffile{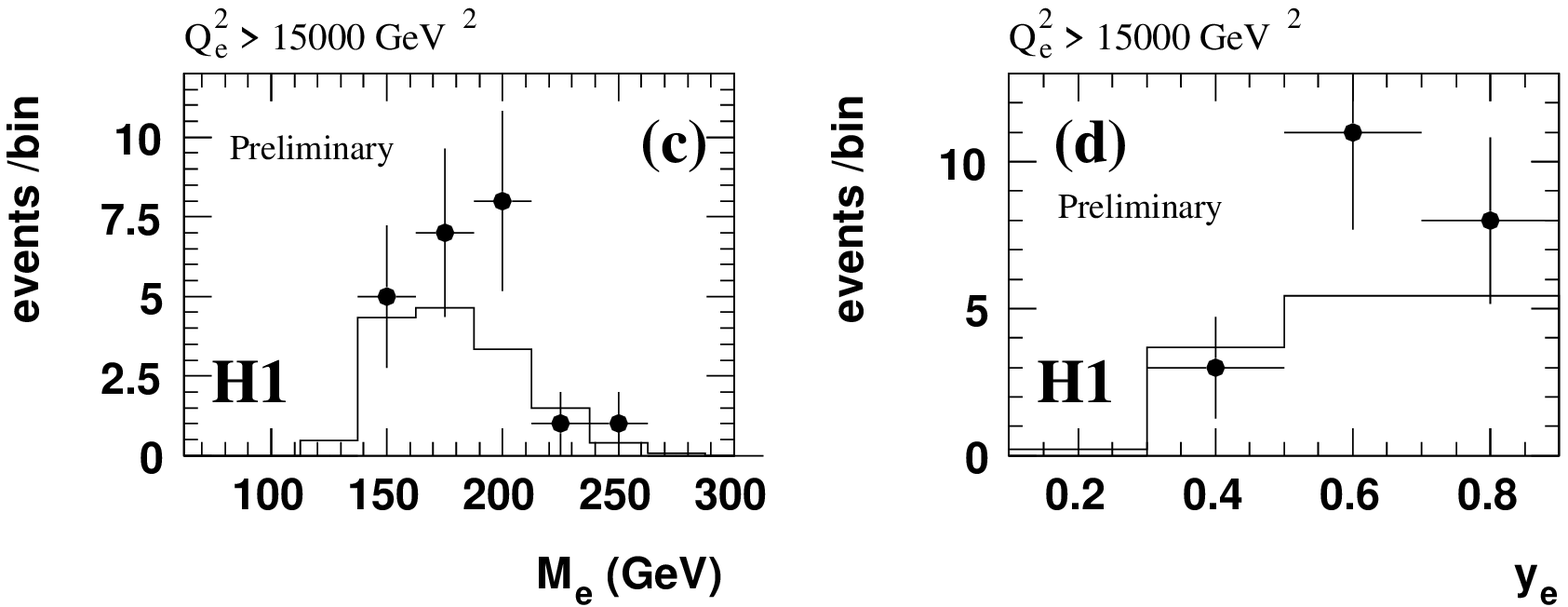}
   \caption[]{ \label{fig:mandy}
     {\small $M_e$ and $y_e$ distributions for H1 NC DIS candidate 
       events ($\bullet$) at $Q^2_e > 15000 \GeV^2$ for (a)(b) the 
       1994$\rightarrow$96 data and (c) (d) the 1994$\rightarrow$97 
       data. Superimposed as open histograms are the standard
       NC DIS expectation.}}
  \end{center}
\end{figure}
% -----------------------------------------------------------------------
For any DIS-like event, as for the production of an $X_S$ particle involved 
in a $2 \rightarrow 2$ body process, a mass $M=\sqrt{s_{ep} x}$ can be 
calculated. 
The Lorentz invariant $x$ represents (at lowest order) the momentum fraction 
of the proton carried by the incident struck quark.
By kinematic constraints, $M$ and $y$ (hence $Q^2$) can be 
reconstructed from two independent measurements such as the energy and angle 
of the final state lepton, or by combining two angles from the lepton 
and hadronic energy flow~\cite{HERAKINE}. 
% --------------- FIGURE : ----------------------------------------------
\begin{figure}[htb]
  \begin{center}
   \epsfxsize=1.0\textwidth 
    \epsffile{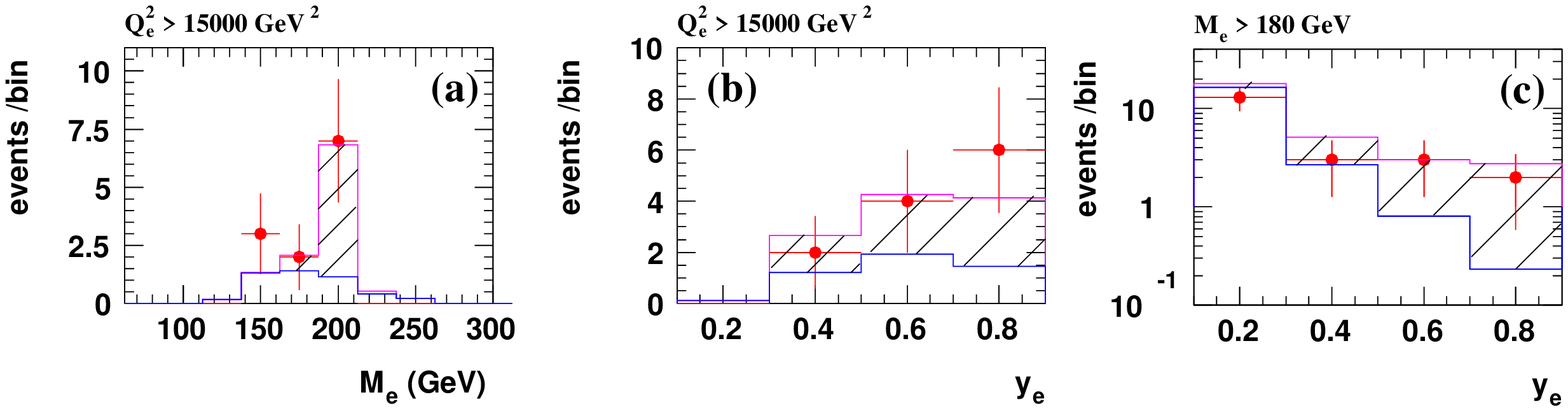}
   \caption[]{ \label{fig:mylego}
       {\small Same distributions as in Fig.~\ref{fig:mandy} taken from
         ref.~\cite{H1HIQ2} but showing in addition to NC DIS, as an 
	 illustration, the expectation for a $M = 200 \GeV$ scalar 
	 leptoquark. }}

  \end{center}
\end{figure}
% ----------------------------------------------------------------------

Fig.~\ref{fig:q2rates}a shows the $Q^2_e$ distribution 
measured~\cite{H1ICHEP533} by H1 for the 1994$\rightarrow$97 data 
in comparison with the SM expectation for NC DIS.
The ratio of the measured over expected distributions is shown
in Fig.~\ref{fig:q2rates}b.
The essential features observed~\cite{H1HIQ2} in the original data remain 
visible.
The NC DIS expectation is seen to be in excellent agreement with 
the data for $Q^2_e \lsim 10000 \GeV^2$.
At larger $Q^2_e$, deviations are observed, with a slight deficit in the
range  $Q^2 \simeq 10000-15000 \GeV^2$ and a number of observed events 
in excess at $Q^2 \gsim 15000 \GeV^2$.
Similar plots are now shown in Fig.~\ref{fig:q2rates}c,d for H1 1997 data  
alone. 
It is seen that the new data ``suggests'' similar deviations but with 
marginal significance.
Considering the full set of $e^+p$ data, H1 and ZEUS are 
left~\cite{H1ICHEP533,ZEUSICHEP} with slight excesses at highest $Q^2$.
For $Q^2 > 15000 \GeV^2$, H1+ZEUS observe $N_{obs}^{H1+ZEUS} = 42$ events 
while $N_{DIS} = 32 \pm 8.5$ events are expected. 
For $Q^2 > 20000 \GeV^2$, $N_{obs}^{H1+ZEUS} = 18$ while 
$N_{DIS} \simeq 9.5 \pm 1$. 
One expects an equal or larger upward fluctuation in $\sim 1 \%$ of random 
experiments.
ZEUS most significant $Q^2$ deviation comes from 2 outstanding events 
at $Q^2_{da} > 35000 \GeV^2$ from their original dataset~\cite{ZEUSHIQ2}, 
where they now expect $0.29 \pm 0.02$.

%
% Rate measurements: Mass and Y distributions
%

Fig.~\ref{fig:mandy}a and b show the H1 $M_e$ and $y_e$ distributions at 
$Q^2_e > 15000 \GeV^2$ for the original 1994$\rightarrow$96 
dataset. 
The data was seen to exceed SM expectation most prominently around 
$M_e \sim 200 \GeV$ for large $y_e$.
Out of the $12$ H1 events at $Q^2 > 15000 \GeV^2$, 7 appeared 
to be ``clustered'' in the bin $ 200 \GeV \pm \Delta M / 2$ with 
$\Delta M = 25 \GeV$ where one would expect $0.95 \pm 0.2$.
This particular clustering was not specifically supported by 
the original ZEUS observations~\cite{DREES}.
I will come back on this question in section~\ref{sec:leptoquark}.
The same representation of H1 $M_e$ and $y_e$ results but including 1997 
data are shown in Fig.~\ref{fig:mandy}c and d.
The excess in the mass bin at $M \simeq 200 \GeV$ is found to 
be less significant.
Here the full dataset was re-analysed including a new {\it in situe}
electron energy calibration~\cite{H1ICHEP579,BRUELPHD}.
This has led to slight migrations (within originally quoted 
systematic errors) of individual events in the $M-y$ plane. 
In particular the $M_e$ values are measured on average to be $2.4\%$ 
higher. Considering nevertheless the same central mass bin of
$ 200 \GeV \pm \Delta M / 2$ with $\Delta M = 25$ where the
most significant fluctuation was originally observed, H1 now finds
$N_{obs} = 8$ events while $N_{DIS} = 3.0 \pm 0.5$ are expected.
Of these observed events, 5 originate from the $1994 \rightarrow 96$ 
data (for $38\%$ of ${\cal{L}}_{H1}$) and 3 from the $1997$ data 
(for $62\%$ of ${\cal{L}}_{H1}$).
Hence, it is fair to say that the ``clustering'' around 
$M_e \sim 200 \GeV$ is not confirmed by the 1997 data alone.

Shown for illustration in Fig.~\ref{fig:mylego} are $M_e$ and $y_e$ 
contributions expected for a scalar leptoquark of mass $M_{X} = 200 \GeV$ 
and given coupling~\cite{BRUELPHD}. 
Such a new particle would only contribute significantly at highest $Q^2$. 
Hence, in presence or not of such new physics, it is reassuring that H1
and ZEUS have found~\cite{H1HIQ2,ZEUSHIQ2} an excellent agreement between 
data and SM expectation for the $y$ or mass spectra when considering 
low minimal $Q^2$ thresholds.
Interestingly, at highest $Q^2$ the shape of the original $M_e$ and $y_e$ 
distributions and the excess are seen to be very well described by a 
combination of NC DIS $+$ leptoquark contributions.

%=======================================================================
\section{Differential Cross-Sections at High $Q^2$ at HERA}
\label{sec:xsect}

%
%  Differential X-sections :
% 
The observed $Q^2_e$ rates can be turned into differential
cross-sections by converting the measured number of events to bin 
averaged values, using Monte Carlo acceptance calculations and 
detector efficiencies. 
% --------------- FIGURE : Differential Cross-section --------------------
\begin{figure}[ht]

  \begin{tabular}{p{0.35\textwidth}p{0.65\textwidth}}
   \caption[]{ \label{fig:dq2plot}
       {\small  Differential cross-section $d\sigma/dQ^2$
                for NC and CC DIS processes measured by the H1 
		(squares) and ZEUS (points) with $e^+p$ 
		data and compared to SM expectation (using here CTEQ4 for
		the parton momentum distributions). }}

 &
   \hspace*{-1.0cm} \raisebox{-150pt}{\mbox{\epsfxsize=0.63\textwidth 
               \epsffile{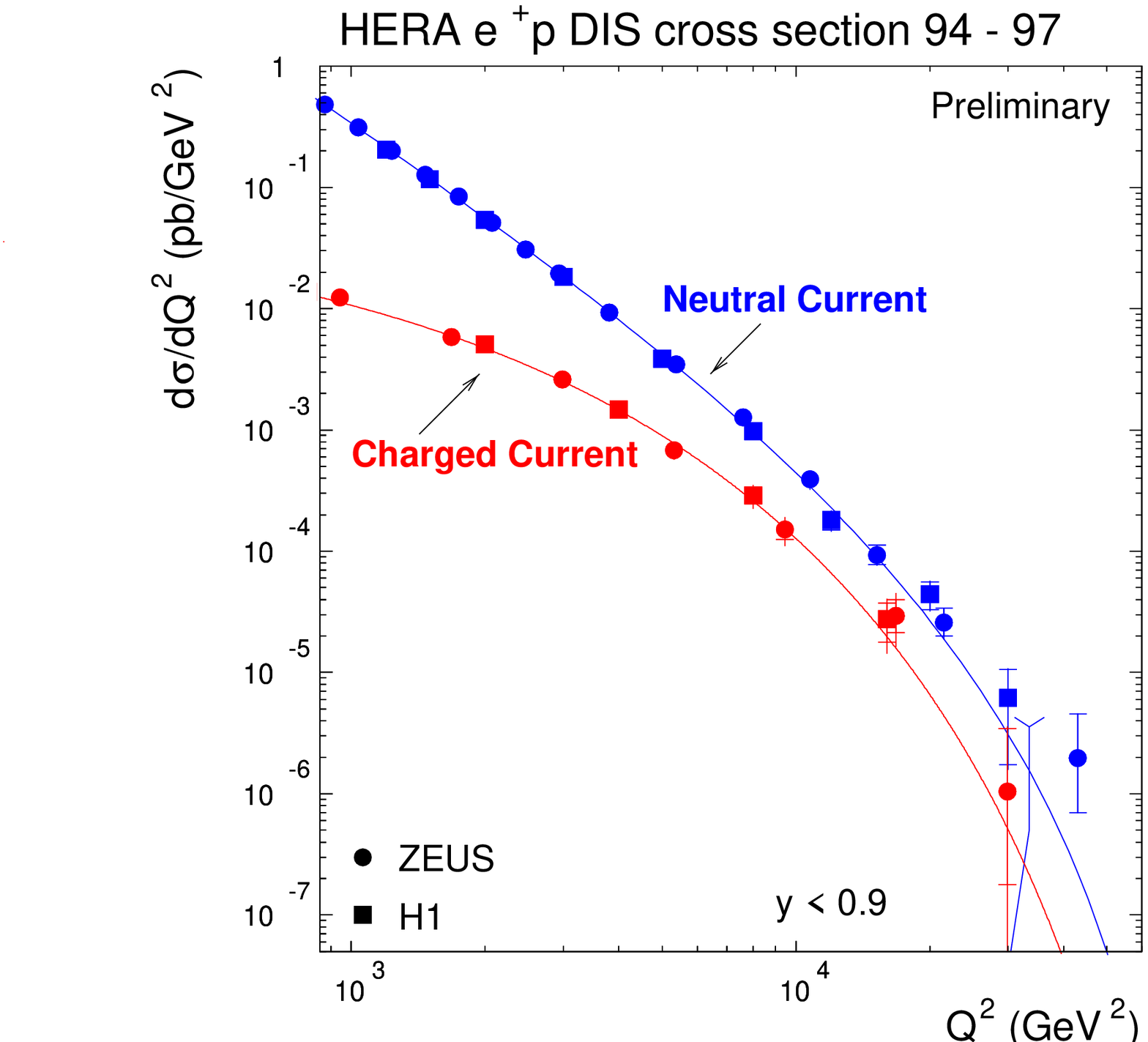}}}
 \end{tabular}
\vspace*{-0.5cm}

\end{figure}
%------------------------------------------------------------------
It should be emphasized that such procedure implicitly assumes that the 
simulation of the SM model properly accounts for migrations (e.g. due to 
initial or final state gluon radiation) from true to measured kinematic 
quantities. This assumption could be invalidated in presence of new 
physics such as resonant production of long-lived leptomesons~\cite{LQMESONS}.
The differential cross-sections $d\sigma/dQ^2$ extracted by the
H1~\cite{H1ICHEP533} and ZEUS experiments~\cite{ZEUSICHEP}
are shown for NC and CC DIS in Fig.~\ref{fig:dq2plot}. 

Both experiments find for NC a remarkable agreement with SM expectation
for $Q^2 \lsim 10000 \GeV^2$ over 4 orders of magnitude.
It should be recalled for completeness that in this $Q^2$ range and at 
large $x$, the SM contribution for NC DIS proceeds dominantly through 
$t$-channel $\gamma$-exchange with a $u$ valence quark of the proton.
For CC DIS, due to the $W$ mass term in the exchanged boson propagator,
the cross-section is suppressed at low $Q^2$, and falls less steeply than
for NC in the intermediate $Q^2$ range.
In $e^+p$, CC proceeds dominantly through $t$-channel exchange
with a $d$ valence quark of the proton.
Thus, the CC cross-section remains a factor $\gsim 4$ below that of NC
for $Q^2 \gsim 10000 \GeV^2$.

\begin{figure}[htb]
  \begin{tabular}{p{0.35\textwidth}p{0.65\textwidth}}
    \caption[]{ \label{fig:dsdq2osm}
    {\small Ratio of the measured over the SM predicted NC $d\sigma/dQ^2$
            cross-section obtained by (a) H1 and (b) ZEUS. 
	    Here for the SM, H1 takes a NLO QCD fit extrapolated 
 	    from $Q^2 \lsim 120 \GeV^2$ data while ZEUS makes use of the
	    CTEQ4D QCD evolved parton densities.
            The gray band shows 
 	    (a) for H1 the uncertainty on the absolute normalisation 
	      (i.e. measured ${\cal{L}}$) and 
	    (b) for ZEUS the uncertainty on parton density  
 	    functions.
 	    Inner (outer) bars correspond to statistical (total) errors. }}
&
      \raisebox{-110pt}{\mbox{\epsfxsize=0.60\textwidth 
                        \epsffile{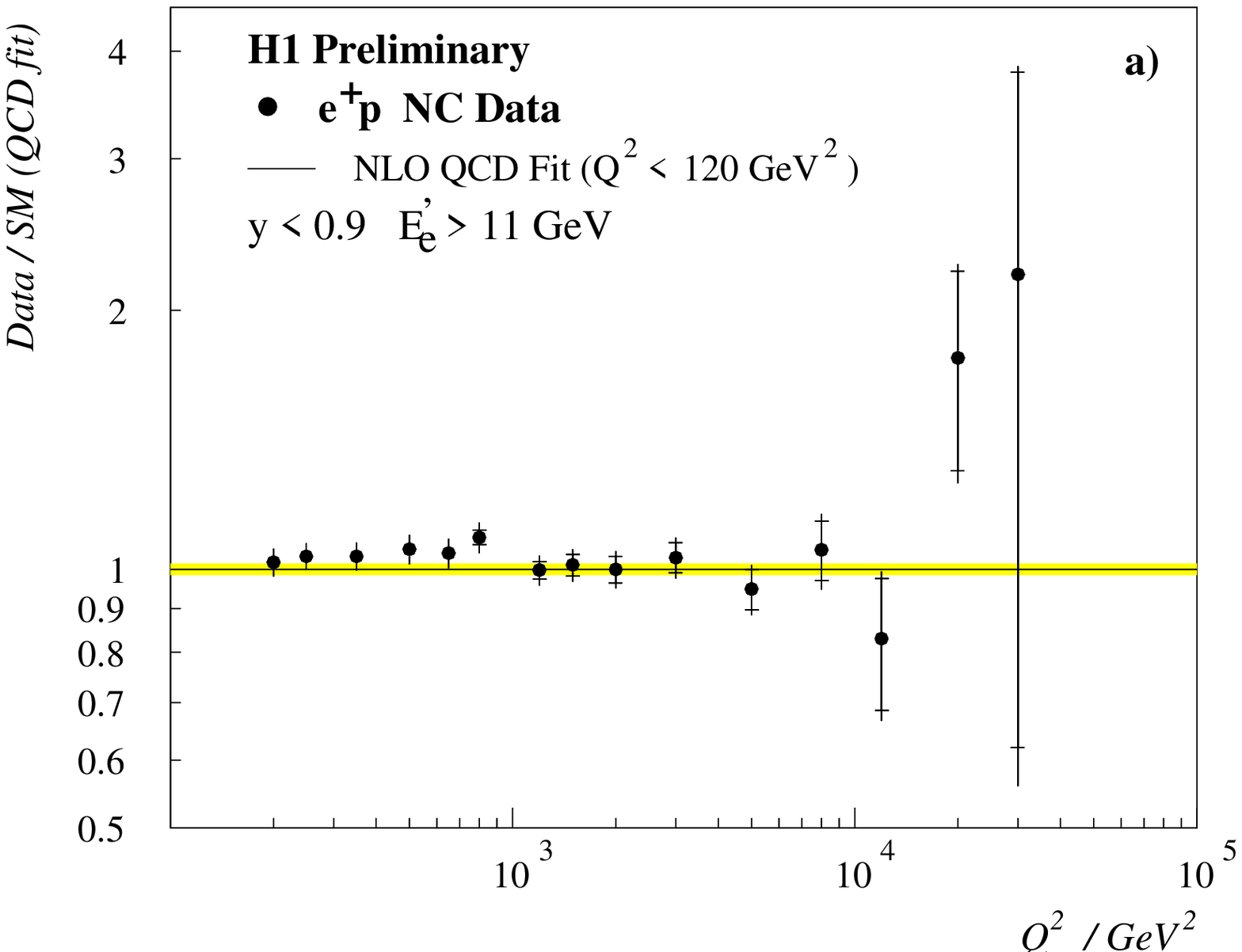}}}
      \hspace*{-7.5cm} \raisebox{-270pt}{\mbox{\epsfxsize=0.62\textwidth 
                        \epsffile{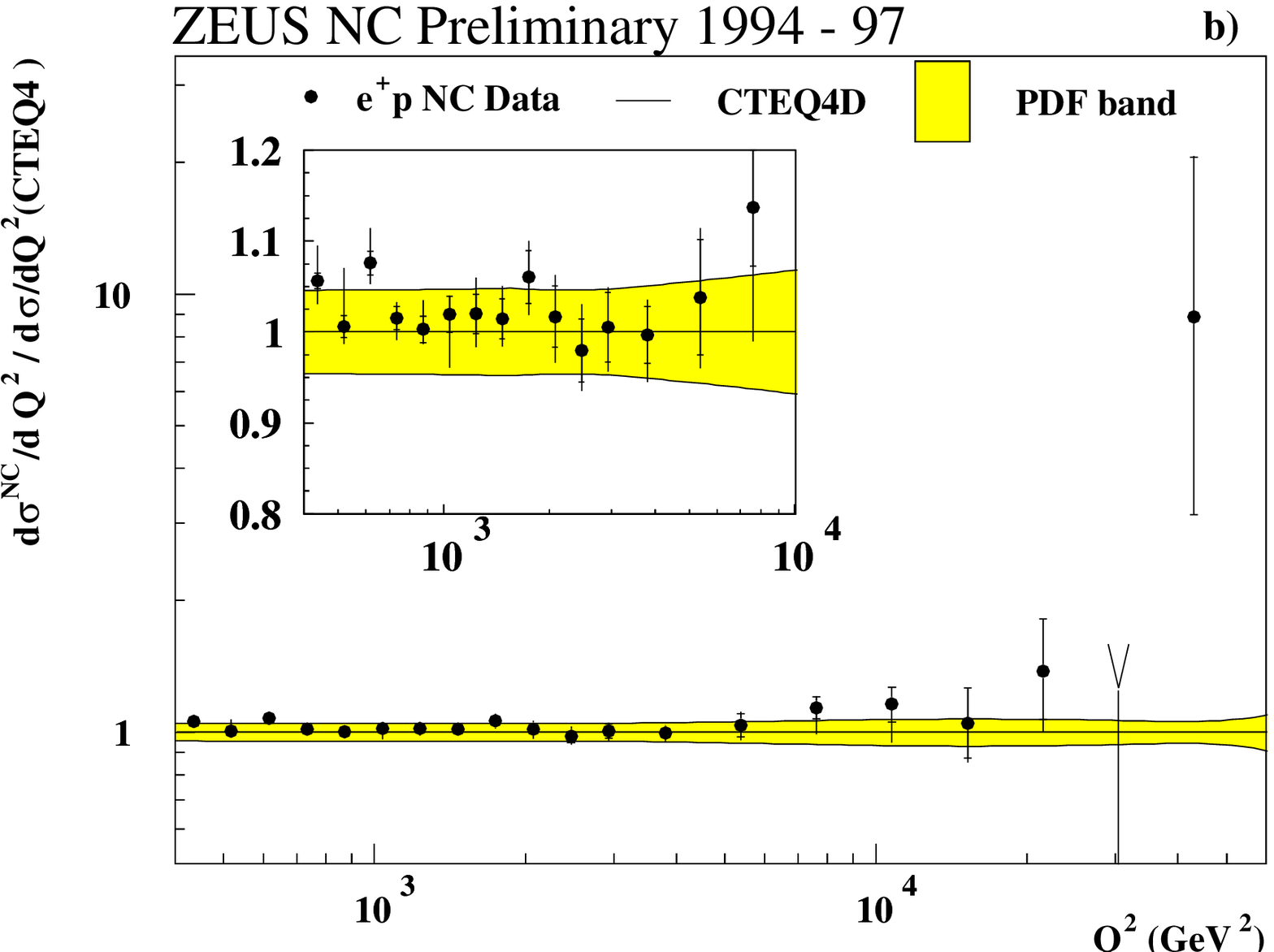}}}
 \end{tabular}
 \vspace*{-0.4cm}

\end{figure}

Fig.~\ref{fig:dsdq2osm}a and b show for H1 and ZEUS the ratio of the measured 
over expected NC $d\sigma/dQ^2$ cross-sections. 
%------------------ FIGURE : -----------------------------------
% 
%-------------------------------------------------------------------------
Here for the SM prediction, H1 performs~\cite{H1ICHEP533} a NLO
QCD fit while ZEUS uses the CTEQ4 parton momentum distributions~\cite{CTEQ}.
To constrain the high $x$ domain, the H1 fit combines the $F_2$ 
structure function data from NMC~\cite{F2NMC} and BCDMS~\cite{F2BCDMS} 
on both proton and deuteron targets. It furthermore makes use of low $Q^2$ 
H1 data ($Q^2 \lsim 120 \GeV^2$) from 1994 and 1995-96~\cite{F2H19496}.
The measurement is clearly seen here to be in excellent agreement
with SM expectation in the $Q^2$ range $1000 < Q^2 < 10000 \GeV^2$.
As was already seen from ``unbiased'' event rates in 
Fig.~\ref{fig:q2rates}, the H1 measurement slightly undershoots the
SM expectation at $Q^2 \sim 10000 \GeV^2$ while an excess is observed
at the $2 \sigma$ level for $Q^2 \gsim 15000 \GeV^2$.
For CC DIS, H1 and ZEUS had observed in their original 
analysis~\cite{H1HIQ2,ZEUSHIQ2} and still 
observe~\cite{ZEUSICHEP,H1ICHEP579,HERAMORIOND} a tendency for the data to lie 
above SM expectation at highest $Q^2$; but the CC results receive large 
systematic error contributions from the experiments (hadronic energy scale) 
and ``model'' ($d$ quark momentum density). 
For a measured $Q_h^2 > 15000 \GeV^2$,
H1 observes $9$ events for $5.1 \pm 2.8$ expected while ZEUS observes 
$8$ events for $3.9^{+1.9}_{-1.6}$ expected.

% --------------- FIGURE : ---------------------------------------------------
\begin{figure}[htb]
\vspace*{-0.4cm}

\begin{center}
\mbox{\epsfig{figure=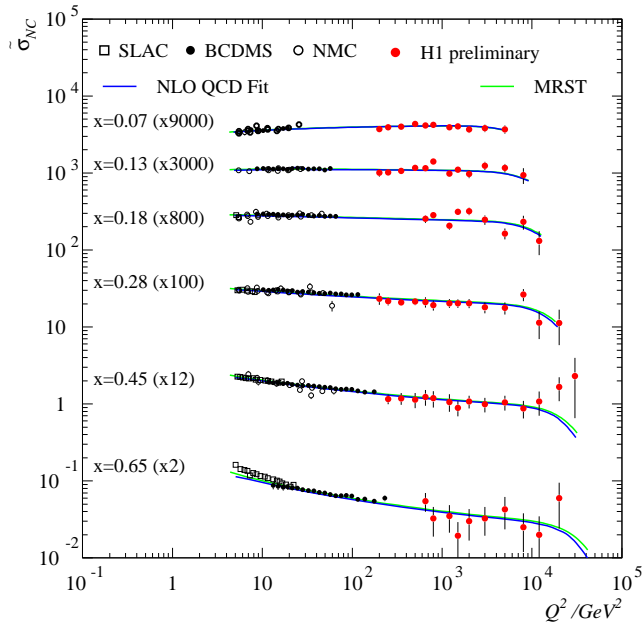,clip=,width=0.70\textwidth,%
bbllx=40pt,bblly=110pt,bburx=560pt,bbury=690pt}}
   \vspace*{-0.5cm}
   
   \caption[]{ \label{fig:dxdq2plot}
      { \small  Measured H1 $d^2\sigma/dxdQ^2$ reduced NC cross-section ($\bullet$)
                compared to the MRST parametrization (upper curves) and to a new
		NLO QCD fit (lower curves) combining H1 preliminary results 
		with BCDMS and NMC.}}
\end{center}
\end{figure}
%-------------------------------------------------------------------------------
The $d^2\sigma/dxdQ^2$ cross-section extracted by H1 is plotted 
as a function of $Q^2$ for a wide range of fix $x$ values in 
Fig.~\ref{fig:dxdq2plot}. 
The results are expressed in 
terms of the reduced cross-section $\tilde{\sigma}_{NC}$ defined as
$\tilde{\sigma}_{NC} \equiv (x Q^4/2\pi\alpha^2) 
                          \, 1/(1+(1-y)^2) \, d^2\sigma/dxdQ^2$.
They are  compared to the MRST structure function fit extrapolated to 
high $Q^2$ as well as to a NLO QCD fit~\cite{H1ICHEP533} taking into 
account the H1 high $Q^2$ data. 
The high $Q^2$ data is seen to further pull the NLO QCD fit
downward at highest $Q^2$ relative to MRST.
The upward ``fluctuation'' discussed above is, here consistently,
visible at $x \simeq 0.45$ and $Q^2 \gsim 15000 \GeV^2$.
From an analysis~\cite{HEINEDIS98} of the $d\sigma/dx$ cross-section for 
$Q^2 > 10000 \GeV^2$, and for the bulk of the cross-section which sits at 
the intermediate $x$ range of $x \simeq 0.2-0.3$, the suppression due to 
the negative $\gamma-Z^0$ interference in $e^+p$ collisions (strikingly 
manifest from the inflexion of the theoretical prediction in 
Fig.~\ref{fig:dxdq2plot}) has now been solidly confirmed by both HERA 
experiments~\cite{ZEUSICHEP,H1ICHEP579}.

%
%=======================================================================
\section{Searches for Contact Interactions}
\label{sec:contact}

Through the interference with SM gauge boson exchange, new bosons of
mass $M \gg \sqrt{s_{ep}}$ could affect the $d\sigma/dQ^2$ cross-section
measurements at HERA. Such a new interaction can be described as an
effective $4$-fermion ``pointlike'' $(\bar{e}e)(\bar{q}q)$ contact
interaction (CI). CI was discussed~\cite{HIGHXYCI} as a possible
``explanation'' of an excess of very high $Q^2$ events at HERA.
%
% --------------- FIGURE : --------------------------------------------
\begin{figure}[htb]
\vspace*{-0.5cm}

  \begin{tabular}{cc}
    \mbox{\epsfxsize=0.5\textwidth 
      \epsffile{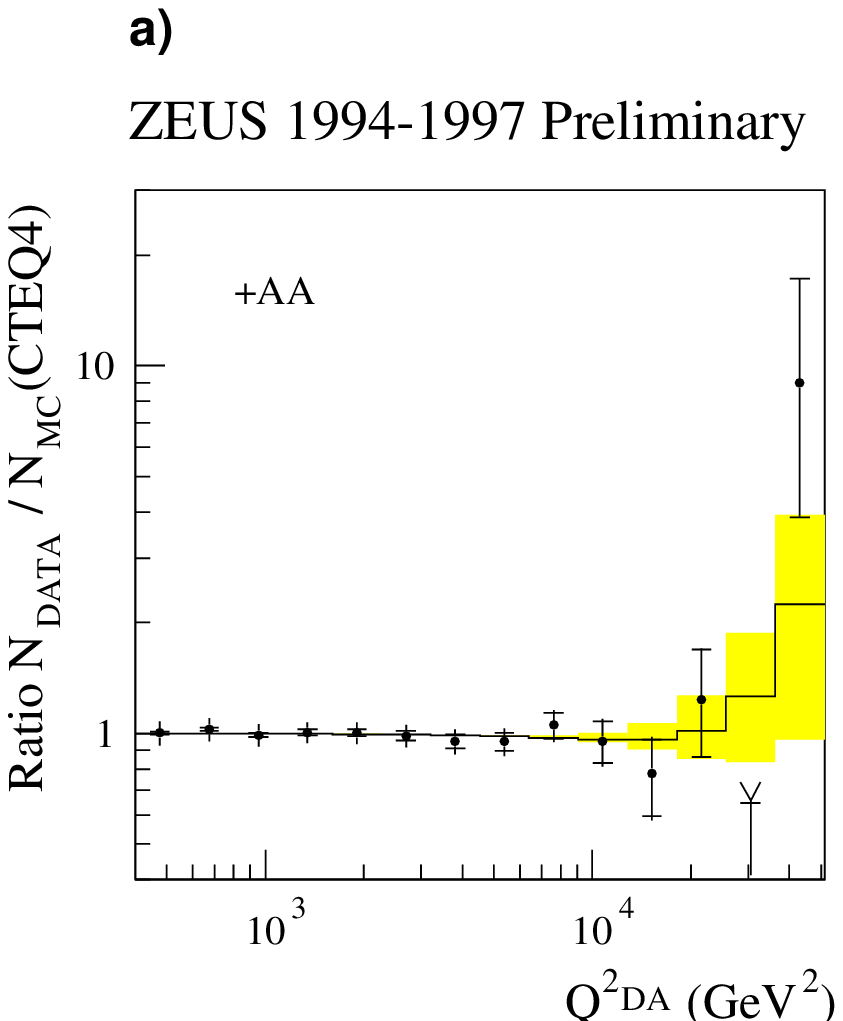}}
   &
    \hspace*{-0.2cm} \mbox{\epsfxsize=0.45\textwidth 
     \epsffile{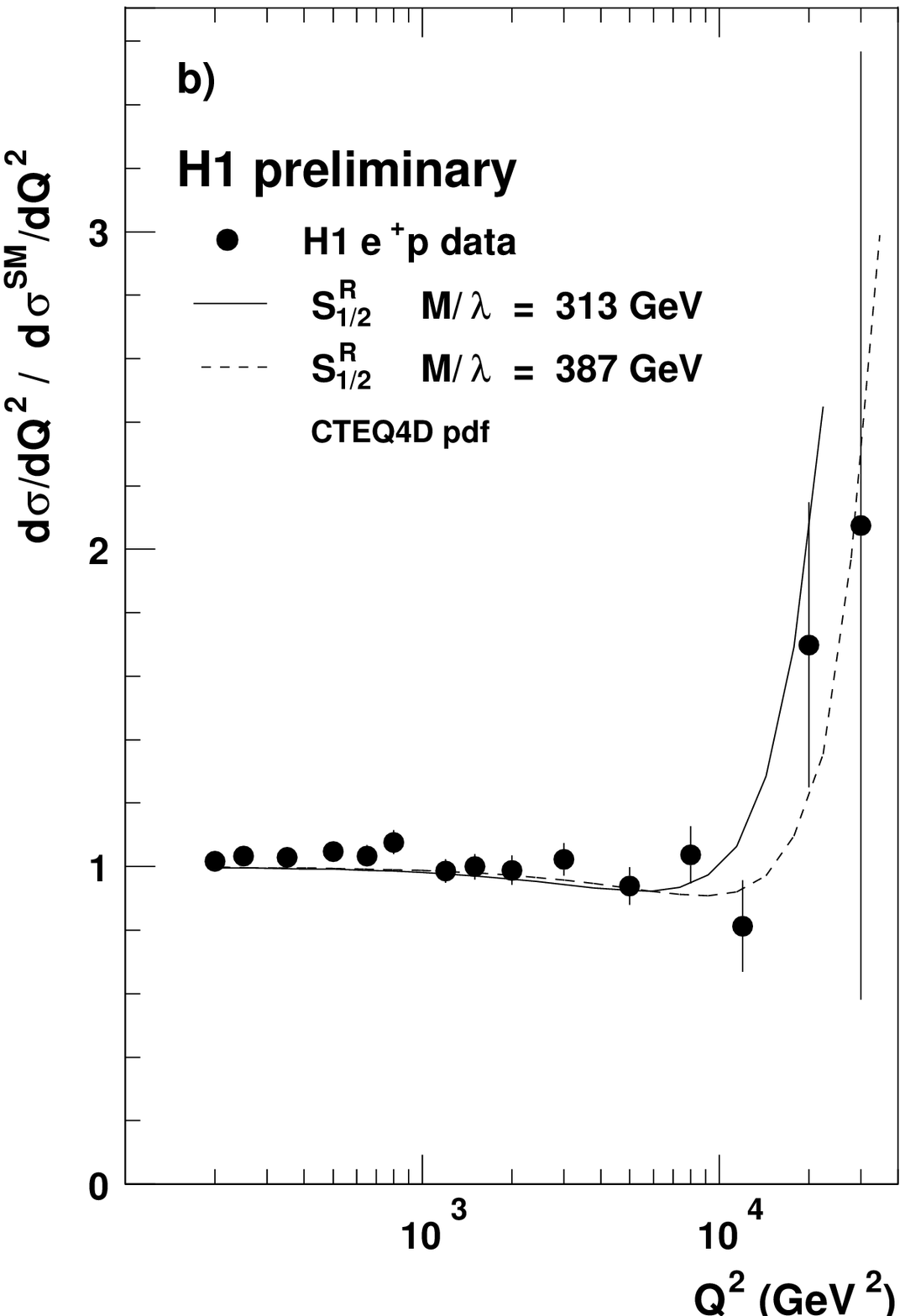}} 
  \end{tabular}
  \caption[]{ \label{fig:dsdq2ci}
      {\small (a) Ratio of the number of observed events per $Q^2$ 
	          interval in ZEUS to the SM expectation. The full 
		  line shows the contact interaction (CI) best fit 
		  for the +AA model (see text); the shaded area is 
		  bounded by the $\pm 1 \sigma$ limits. 
              (b) Ratio of the measured $d\sigma/dQ^2$ to the SM 
	          expectation for NC-like events in H1; the full line 
		  (dashed line) gives for a CI the 95\% CL (resp. the 
		  CI best fit) for the $S_{1/2}^{R}$. }}
\end{figure}
% ---------------------------------------------------------------------
The sensitivity to CI has now been investigated~\cite{HERAICHEPCI} using 
all available $e^+p$ data and considering the interaction Lagrangian 
${\cal{L}}_{int} = {\cal{L}}_{SM}^{NC} + {\cal{L}}_{CI}^{NC}$
where the chiral invariant ${\cal{L}}_{CI}^{NC}$ complementing the
SM can be written in the form 
${\cal{L}}_{CI}^{NC} = \sum \left\{ 
  \eta^q_{ij} (\bar{e}_i \gamma_{\mu} e_i)(\bar{q}_j \gamma^{\mu} q_j)
                                           \right\} $
with $q=u,d$ and $i,j=L,R$ and the coupling coefficients defined 
as $\eta_{ij} \equiv \pm g^2 / \Lambda_{ij}^{\pm 2}$. 
The $+$ ($-$) sign indicate constructive (destructive) interference.
The ``mass scale'' $\Lambda^{\pm}$ is conventionally defined when
setting constraints as that scale relevant for a ``strong'' coupling 
strength of $g^2 = 4\pi$.

The chirality structure of the CI model can be chosen to avoid the severe
constraints coming from atomic parity violation~\cite{APVIOL} (APV).
These are cancelled in particular if for the quarks $q$,
$\eta^q_{LL} + \eta^q_{LR} - \eta^q_{RL}+ \eta^q_{RR} = 0$, as 
realized for instance in the $VV$, $AA$ and $VA$ models considered
by H1 and ZEUS, with the mixture
$VV = LL + LR + RL + RR$, $VA = LL - LR + RL - RR$, and
$AA = LL - LR - RL + RR$.
The SU(2) invariance is assumed which imposes $\eta^u_{LL} = \eta^d_{LL}$ 
and $\eta^u_{RL} = \eta^d_{RL}$.
%
% --- TABLE : Contact Interaction limits  -----------------------
%
\begin{table*}[htb]
  \renewcommand{\doublerulesep}{0.4pt}
  \renewcommand{\arraystretch}{1.2}
 \begin{center}
 \vspace*{-0.1cm}
 
 \begin{tabular}{p{0.30\textwidth}p{0.70\textwidth}}
         \caption
         {\small \label{tab:cilimits}
	   Constraints at 95\% CL on contact interaction scale 
	   $\Lambda$ from HERA and other colliders in models 
	   with VV, AA or VA structure (see text) for constructive
	   ($\Lambda^+$) or destructive ($\Lambda^-$) interference.}
&
 \vspace*{-0.1cm}

   \begin{tabular}{||c|c||c|c|c|c||}
   \hline \hline
   \multicolumn{2}{||c||}{Model} & \multicolumn{4}{c||}{95\% CL Lower Limits} \\  
   \multicolumn{2}{||c||}{ } & 
               \multicolumn{2}{c|}{HERA} & TeV & LEP  \\  
   \multicolumn{2}{||c||}{  } & H1 & ZEUS & CDF/D0 & ADLO \\  
   \hline
   VV  & $\Lambda^+$  &\, 4.5 \,& 4.9 & 3.5 & 4.0 \\
       & $\Lambda^-$  &\, 2.5 \,& 4.6 & 5.2 & 5.2 \\
   \hline
   AA  & $\Lambda^+$  &\, 2.0 \,& 2.0 & 3.8 & 5.6 \\ 
       & $\Lambda^-$  &\, 3.8 \,& 4.0 & 4.8 & 3.7 \\  
   \hline
   VA  & $\Lambda^+$  &\, 2.6 \,& 2.8 &     &     \\ 
       & $\Lambda^-$  &\, 2.8 \,& 2.8 &     &     \\  
   \hline \hline
   \end{tabular}
 \end{tabular}
\end{center}
\end{table*}
%.......................................................................

Figure~\ref{fig:dsdq2ci}a shows CI constraints and best fit to the $Q^2$ 
spectrum obtained by ZEUS for the $AA (\Lambda^+)$ scenario.
Figure~\ref{fig:dsdq2ci}b shows as an illustration the constraints 
and best fit obtained by H1 in a leptoquark $S_{1/2}^R$ scenario. 
The $S_{1/2}^R$ (see nomenclature in next section) possesses
a $-5/3$ charge state coupling to $e^-_R \bar{u}$ ($e^+_L u$)  and
a $-2/3$ charge state coupling to $e^-_R \bar{d}$ ($e^+_L d$).
Assuming SU(2) invariance, the CI couplings are
$\eta_{RL}^u = \eta_{RL}^d = -1/2 (\lambda /M_S)^2$. 
Both experiment rightly conclude~\cite{HERAICHEPCI} independently that 
no significant indication of a CI was found for these and various other 
models considered~\cite{HERAICHEPCI}.
It is nevertheless interesting to note that the best fits in the 
$AA (\Lambda^+)$ scenario are found, consistently, to allow for a 
non-vanishing coupling.
ZEUS finds $\eta/4\pi = 1/\Lambda_0^2 = 0.16^{+.05}_{-.06}$ while H1
finds $0.15^{+.04}_{-.07}$ ; corresponding roughly to a value of
$\Lambda_0 \simeq 2.5 \TeV$.
In practice, such a simple CI scenario is actually ruled out by TeVatron
and LEP constraints as seen in Table~\ref{tab:cilimits}.
It is also interesting to note that, as seen in Fig.~\ref{fig:dsdq2ci}b
an hypothetical scalar can be accommodated by H1 data for 
$M_S/\lambda = 387 \GeV$.
For the $VV$ and $VA$ scenario, the HERA sensitivity is found to be
comparable to that of other colliders.
The $VV(\Lambda^-)$ scenario would lead to a suppression of the NC
$d\sigma/dQ^2$ up to a few $10^4 \GeV^2$ followed by a strong 
enhancement~\cite{HERAICHEPCI} qualitatively similar to the actual 
observation.
 
%
%=======================================================================
\section{Searches for Leptoquarks and Lepton Flavour Violation}
\label{sec:leptoquark}
 
%
% Leptoquark Phenomenology
% %%%%%%%%%%%%%%%%%%%%%%%%
%

Generic leptoquarks (LQ) are colour triplet bosons which appear naturally 
in various unifying theories beyond the SM such as Grand 
Unified Theories and Superstring inspired $E_6$ models, and in some 
Compositeness and Technicolour models. 
LQ searches have been carried either in the strict context of the 
original Buchm\"uller-R\"uckl-Wyler (BRW) effective model~\cite{BUCH1987} 
where the decay branching ratios are fixed by the model, or in the context 
of generic models allowing for arbitrary branching ratios.  

The BRW model considers all possible scalar ($S_I$) and vector ($V_I$) 
LQs of weak isospin {\it I} with dimensionless Yukawa couplings
$\lambda^{L,R}_{ij}$ to lepton-quark pairs, where {\it i} and {\it j}
indices denote lepton and quark generations respectively and
{\it L} or {\it R} is the chirality of the lepton.
The general effective Lagrangian obeys the symmetries of the SM
and introduces 10 different LQ isospin multiplets, among which 5 are 
scalar families. 
These are listed in Table~\ref{tab:brwscalar} in the so-called Aachen 
nomenclature and classification scheme~\cite{LQAACHEN}.
The LQ search can be restricted to pure chiral couplings of the LQs 
given that deviations from lepton universality in helicity suppressed 
pseudoscalar meson decays have not been observed~\cite{DAVIDSON}. 
This restriction to couplings with either left- ($\lambda^L$) or
right-handed ($\lambda^R$) leptons 
(i.e. $\lambda^L \cdot \lambda^R \sim 0$),
affects only two scalar LQs ($S_0$ and $S_{1/2}$). 
In all the results presented below, it is implicitly assumed as a
simplifying assumptions that one of the LQ isospin doublet or triplet
is produced dominantly and that the mass eigenstates within this
multiplet are degenerate.

%----------------------------------------------------
\subsection{Search for first generation leptoquarks }
\label{sec:leptoq}

%
% Leptoquark Phenomenology
% %%%%%%%%%%%%%%%%%%%%%%%%
%

The search for first generation LQs at HERA involves the 
analysis of DIS-like events at very high $Q^2$. 
The production cross-section $\sigma_{LQ}$ depends on the quark momentum 
density in the proton and approximately scales with $\lambda^2$.
The scalar resonance which can then decay into $e+q$ or $\nu + q$ 
is expected to have a very narrow intrinsic width 
$\Gamma = \lambda^2 M/16\pi$ and the decays into a lepton and a quark jet 
lead to event signatures indistinguishable from SM DIS.
% ------------------ TABLE : Scalar Leptoquarks  -------------------------
\begin{table*}[tb]
  \renewcommand{\doublerulesep}{0.4pt}
  \renewcommand{\arraystretch}{1.2}
 \vspace{-0.1cm}

\begin{center}
    \begin{tabular}{|c|c|c|c|c|c|}
      \hline
      $F=-2$ & Prod./decay & ${\cal B}(e^+q)$
             & $F=0$ & Prod./decay & ${\cal B}(e^+q)$  \\
      \hline
    $^{-1/3}S^\ast_0$     & $e^+_R \bar{u}_R\rightarrow e^+ \bar{u}$ & $1/2$
  & $^{-5/3}S^\ast_{1/2}$ & $e^+_R u_R \rightarrow e^+ u$            & $1$  \\
                          & $e^+_L \bar{u}_L\rightarrow e^+ \bar{u}$ & $1$
  &                       & $e^+_L u_L \rightarrow e^+ u$            & $1$ \\
      \cline{1-3}
      $^{-4/3}\tilde{S}^\ast_0$
        & $e^+_L \bar{d}_L\rightarrow e^+ \bar{d}$ & $1$
  & $^{-2/3}S^\ast_{1/2}$ & $e^+_L d_L \rightarrow e^+ d$            & $1$ \\
      \hline
      $^{-4/3}S^\ast_1$
        & $e^+_R \bar{d}_R \rightarrow e^+ \bar{d}$
         & $1$
  & $^{-2/3}\tilde{S}^\ast_{1/2}$ & $e^+_R d_R \rightarrow e^+ d$ & $1$ \\
      $^{-1/3}S^\ast_1$
        & $e^+_R \bar{u}_R \rightarrow e^+ \bar{u}$
         & $1/2$
             & & &  \\
      \hline
      \hline
    \end{tabular}
          \caption {\small \label{tab:brwscalar}
                   Scalar leptoquarks isospin families in the 
                   Buchm\"uller-R\"uckl-Wyler model.
                   These LQ will be in the following indexed with
                   the chirality of the incoming {\it{electron}} 
                   which could allow their production, e.g.
                   the $\tilde{S}^\ast_0$ will be denoted by
                   $\tilde{S}^\ast_{0,R}$. }
\end{center}
\end{table*}
% ------------------------------------------------------------------------
A characteristic statistical signal of the direct production of LQs in the 
$s$-channel would be a peak in the reconstructed mass distribution.
The LQ mass can be reconstructed from the decay products 
(e.g. a charged lepton and a jet) or from kinematic constraints as 
$M=\sqrt{s_{ep} x}$. 
The isotropic decay of the LQs in their CM frame leads 
to a flat $y$ spectrum. 
In this frame $y=\frac{1}{2}\left(1+\cos{\theta^*}\right)$ with $\theta^*$ 
being the decay polar angle of the final state lepton.
This is markedly different from the
${\rm d} \sigma\,/\, {\rm d} y\sim\,y^{-2}$ 
distribution expected at fixed $x$ for the dominant $t$-channel
$\gamma$-exchange in NC DIS events.
Hence, the signal of first generation LQs would be most prominent 
at high $y$ or equivalently at large $Q^2 = M^2 y$.

\begin{figure}[htb]
  \vspace*{-0.3cm}

  \begin{center}
  \begin{tabular}{cc}
     \mbox{\epsfxsize=0.44\textwidth 
      \epsffile{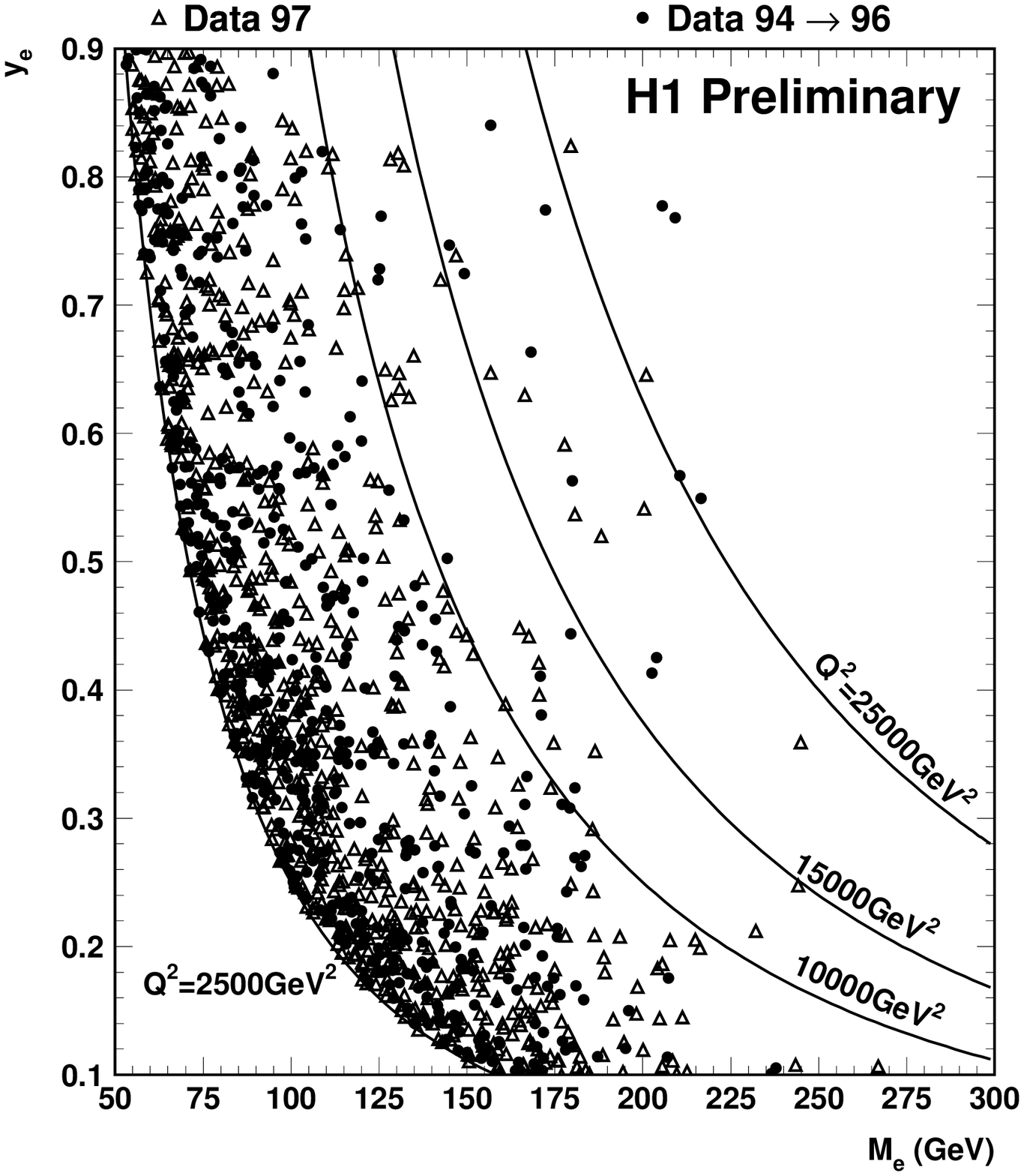}}
   &
     \mbox{\epsfxsize=0.44\textwidth 
      \epsffile{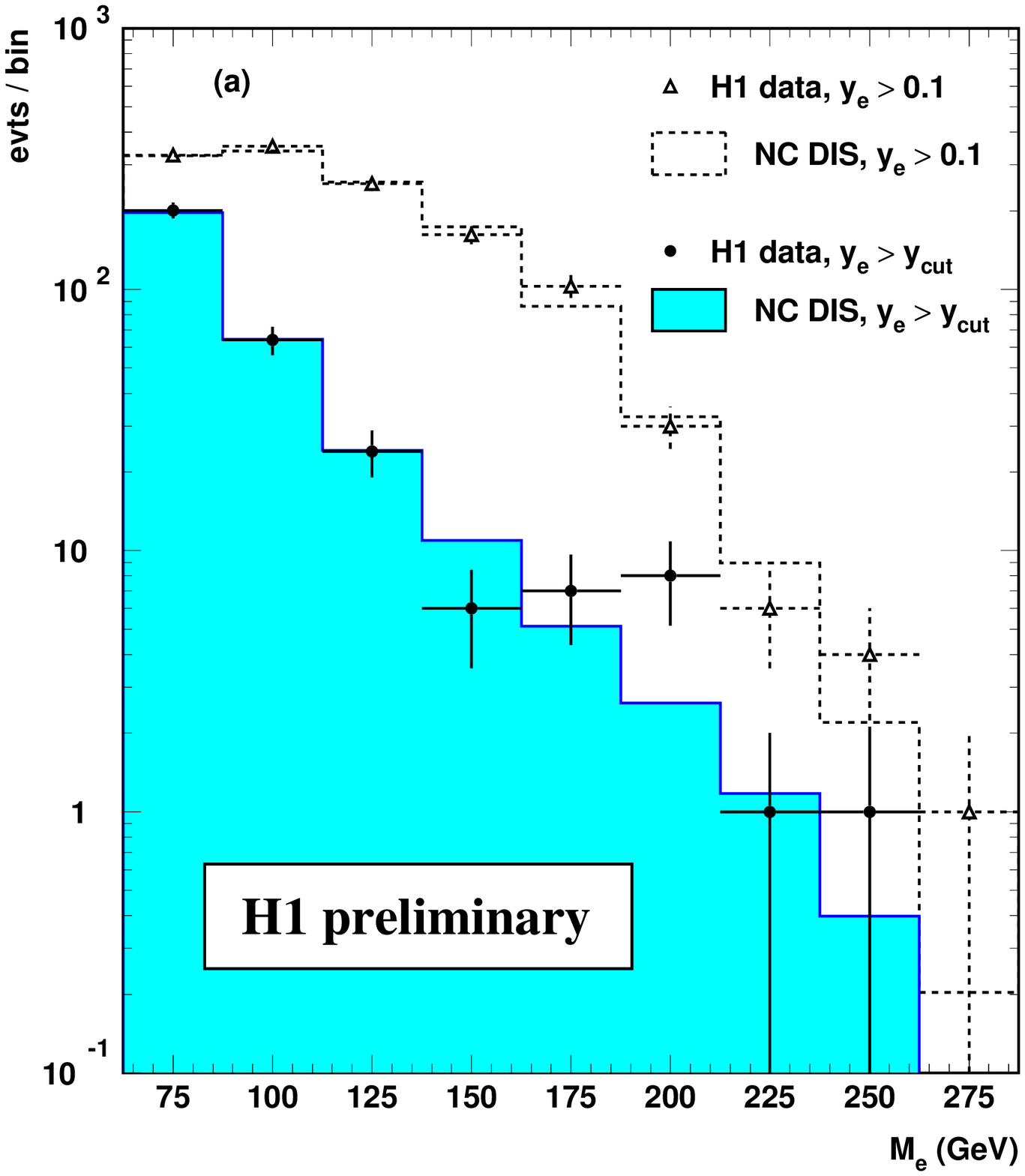}} \\  
     \raisebox{-15pt}{
     \mbox{\epsfxsize=0.44\textwidth 
      \epsffile{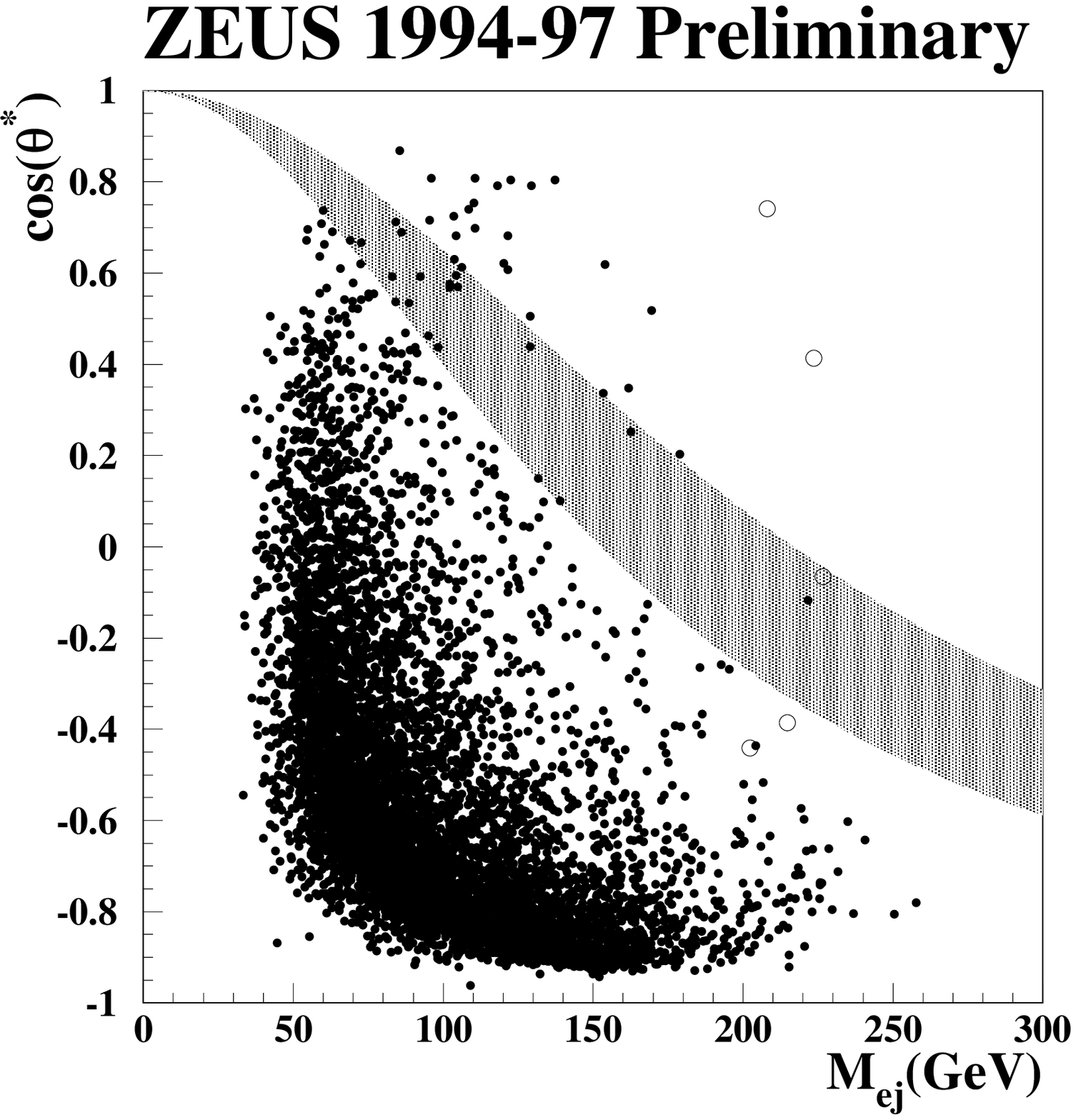}}}
   &
     \raisebox{-15pt}{
     \mbox{\epsfxsize=0.44\textwidth 
      \epsffile{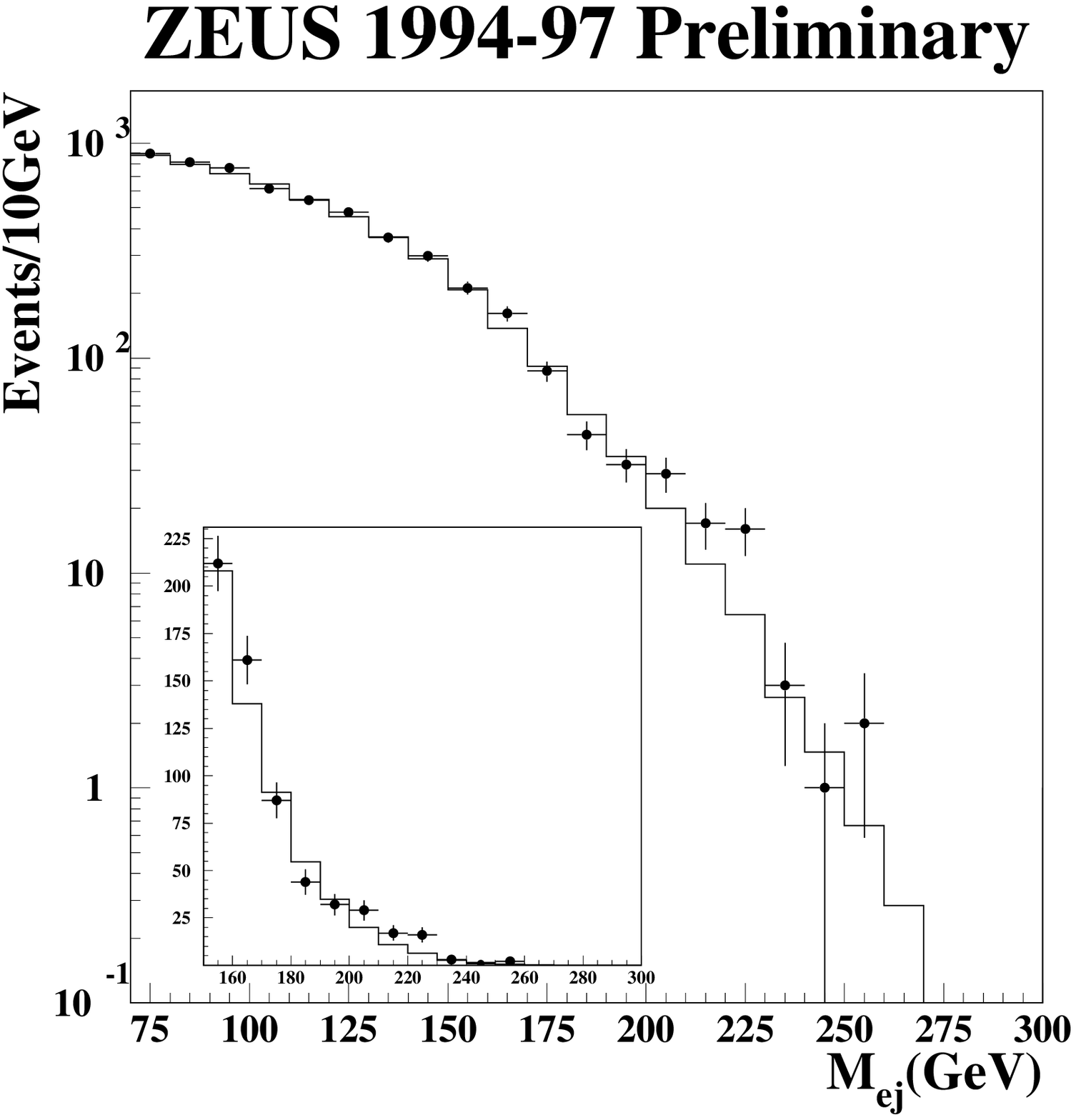}}} \\ 
  \end{tabular}
  \end{center}
     \vspace*{-0.3cm}

 \caption[]{ \label{fig:dndmlq}
 {\small NC DIS candidate events in the $y$ - $M$ plane ({\it left}) and
         mass spectra ({\it right}) from H1 ({\it top}) and 
	 ZEUS ({\it bottom}). }}
\end{figure}

%
% Specific Searches for a Resonance
% %%%%%%%%%%%%%%%%%%%%%%%%%%%%%%%%%
%
% H1
%
The H1 LQ search for NC-like events requires an isolated $e^+$
with a transverse energy of $E_{T,e} > 15 \GeV$. 
The reconstruction of the kinematics relies essentially (except for
about $\simeq 4.4\%$ of azimuthal range) on the $e^+$ energy and
angle. 
The mass and $y$ values of all events with $Q^2 > 2500 \GeV^2$ are shown 
in Fig.~\ref{fig:dndmlq}({\it top left}).
The corresponding measured mass spectrum shape is found to be very well 
predicted by the SM as can be seen in Fig.~\ref{fig:dndmlq}({\it top right}).
Also shown there are the mass spectra after having applied a mass dependent 
$y$ cut~\cite{H1ICHEP533} designed via Monte Carlo studies to optimize the 
signal significance for scalar LQ searches.
This cut varies from $\simeq 0.6$ at $M_e \simeq 60 \GeV$ to $\simeq 0.4$ 
at $M_e \simeq 200 \GeV$, and down to $y_e \simeq 0.2$ at 
$M_e \gsim 250 \GeV$. 
H1 observes 312 events satisfying this $y$ cut in the mass 
range $M_e > 62.5 \GeV$, in excellent agreement with the SM expectation
of $306 \pm 23$.
%
% ---------- FIGURE 10: y vs M + dNdM  NC channel      -----
%
%---------------------------------------------------------------------------
In the mass range $ 200 \GeV \pm \Delta M / 2$ with $\Delta M = 25 \GeV$,
$N_{obs}= 8$ events are found for $N_{DIS} = 3.0 \pm 0.5$. 
As was discussed in section~\ref{sec:highq2}, this slight excess mostly 
originates from the 1994$\rightarrow$96 dataset.
These 8 events have an average mass of $M_e \simeq 206 \GeV$. 
When discussing sensitivity and constraints on LQ searches, it
should be kept in mind that according to LQ Monte Carlo simulations, 
the mass $M_e$ as deduced from the positron tends to systematically underestimate 
a true LQ mass by $\sim 2\%$ for $M_{LQ}$ of ${\cal {O}}(200 \GeV)$.
The most significant fluctuation is observed in H1 for $M_e$ values which would 
correspond to $M_{LQ} \simeq 210 \GeV$.

%
% ZEUS
%

The ZEUS LQ search for NC-like events requires a total $E_T \ge 60 \GeV$. 
A mass $M_{eJ}$ is attributed to each candidate event and calculated from 
the positron and the jet at highest $E_T$, not correcting for the finite
jet mass. 
The $M_{eJ}$ and $\cos \theta^*$ for all events are given in 
Fig.~\ref{fig:dndmlq}({\it bottom left}).
Interestingly, the outstanding high $Q^2$ events already discussed in 
ref.~\cite{ZEUSHIQ2} here tend to cluster around $M_{eJ} \simeq 215 \GeV$. 
The projected $M_{eJ}$ spectrum is shown in Fig.~\ref{fig:dndmlq}({\it bottom right}) 
after having applied conservatively a fiducial cut (shaded area in 
Fig.~\ref{fig:dndmlq}({\it bottom left})) which removes a difficult region 
for electron energy measurements at the interface between two calorimeters, 
thus cutting away 2 very high $Q^2$ events. 
In the remaining acceptance and for $M_{eJ} > 200 \GeV$ and 
$\cos \theta^* > 0.25$, ZEUS observes $7$ events in slight excess 
of the SM expectation of $4.3$ events. Of these $7$ events, $5$ originate 
from the 1994$\rightarrow$96 dataset.
For $M_{eJ} > 200 \GeV$ and in the full $\cos \theta^*$ range, 68 events are 
observed for an expectation of $43^{+14}_{-12}$ from SM.
This excess is found by ZEUS~\cite{ZEUSICHEP} to be mostly due 
to events having a $\cos \theta^*$ spectrum shape compatible with 
SM DIS expectation. 

While interesting upward fluctuations of the number of observed events
have been found by both H1 and ZEUS for masses $M \gsim 200 \GeV$ in their 
% ----------------- FIGURE xx:  Limits lambda, BRW model -----------------
% 
\begin{figure}[htb]
  \begin{tabular}{p{0.35\textwidth}p{0.65\textwidth}}

      \caption[]{ \label{fig:lqxlim}
      {\small Cross-section exclusion limits at $95 \%$ CL from ZEUS for
              generic scalar and vector leptoquarks. }}
 &
      \raisebox{-150pt}{\mbox{\epsfxsize=0.58\textwidth 
      \epsffile{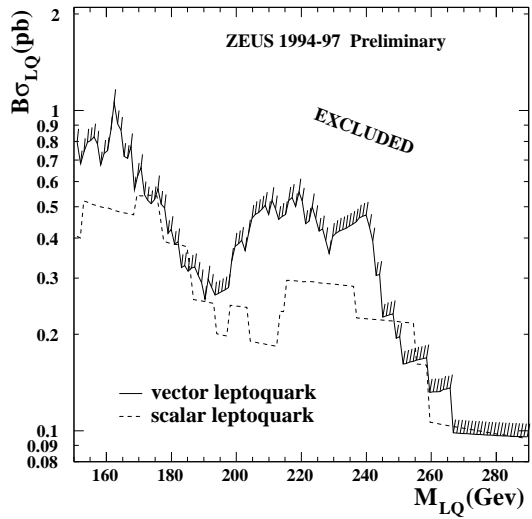}}}
 \end{tabular}
\end{figure}
%-------------------------------------------------------------------------
1994$\rightarrow$96 dataset, the 1997 data disappointingly offered no
confirmation of an excess in the NC DIS-like channel which would possess 
the characteristics of lepton-quark resonant production. 
Hence, both experiments proceed to set mass dependent constraints on the 
production cross-section $\sigma_{LQ}$ of first generation scalar LQs, 
treating the deviations observed as statistical fluctuations.
Model ``independent'' upper limits on $\sigma_{LQ}$ were derived by 
ZEUS in ref.~\cite{ZEUSICHEP} and are shown in Fig.~\ref{fig:lqxlim}.

%
% Exclusion limits from HERA
%
Further interpretation of such constraints can be discussed in the framework 
of the BRW model~\cite{BUCH1987} where $\beta_{eq}$ values are fixed 
by the model for the various LQ types, as given e.g. for scalars in
Table~\ref{tab:brwscalar}.
% ----------------- FIGURE xx:  Limits lambda, BRW model -------------------
% 
\begin{figure}[t]
  \begin{tabular}{p{0.30\textwidth}p{0.70\textwidth}}
     \caption[]{ \label{fig:limilambda}
      {\small Exclusion limits at $95 \%$ CL on the Yukawa coupling
              $\lambda$ as a function of the leptoquark mass for
              $F=0$ scalar leptoquarks described by the BRW model.
              Domains above the curves are excluded. }}
&
     \hspace*{-0.5cm} \raisebox{-120pt}{\mbox{\epsfxsize=0.68\textwidth
              \epsffile{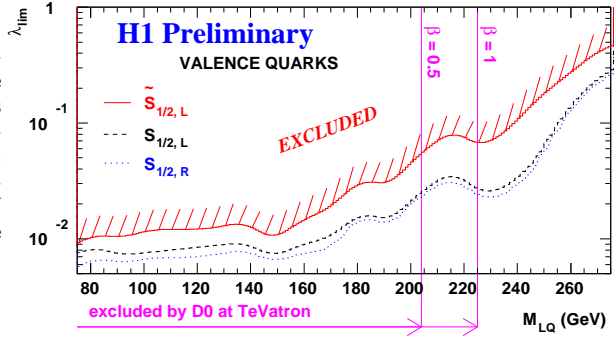}}}
\end{tabular}
\end{figure}
%---------------------------------------------------------------------------
The exclusion upper limits on $\sigma_{LQ}$ translated into mass dependent 
limits on $\lambda_{11}$ as obtained by H1 are shown in 
Fig.~\ref{fig:limilambda} for scalar LQs having fermionic number $F=0$, 
i.e. which can be produced via a fusion between the $e^+$ and a $u$ or 
$d$ valence quark.
The limits are given here for masses up to $275 \GeV$ above which one 
needs to move away from a resonance-like search since the mass peak 
of $F=0$ scalars becomes severely distorted.
Very similar results are obtained by ZEUS who also derived constraints
for vector LQs~\cite{ZEUSICHEP}.
The $e^+ p$ collisions naturally offers the best sensitivity to 
$F=0$ LQs and the limits obtained represent an improvement by a 
factor $\simeq 3$ compared to previously published HERA 
results~\cite{H1ZEUSLQ}.
For a coupling of the electromagnetic strength
$\lambda^2 / 4 \pi = \alpha_{EM}$ (i.e. $\lambda \simeq 0.3$),
such LQ are excluded at $95 \%$ CL up to $275 \GeV$.
The highest sensitivity (hence most severe constraints) is obtained
for $S_{1/2}^R$ since both charge states can be produced via a 
$e^+ u$ and $e^+ d$ fusion. 
Only $e^+ u$ ($e^+ d$) fusion is possible for the $s$-channel 
production of $S_{1/2}^L$ ($\tilde{S}_{1/2}^L$).
The $F=2$ LQs will be best probed with the forthcoming 
$e^-p$ data taking starting in 1998 at HERA. 

The mass range of interest for a possible discovery at HERA of
a LQ of the strict BRW model has now been severely reduced by the 
TeVatron $p \bar{p}$ experiments~\cite{D01GENE,CDF1GENE}, where first 
generation scalar LQs with $M < 242 \GeV$ and $\beta_{eq} =1$ are excluded 
(95\% CL) independently of the $\lambda$ (see ref.~\cite{TEVCOMBINE}).
For $\beta_{eq} = 0.5$, the excluded domain reaches $M \simeq 200 \GeV$.
The mass range excluded by the D0 experiment alone is shown in 
Fig.~\ref{fig:limilambda}.
At LEP $e^+ e^-$ collider, searches~\cite{DELPHI} for single LQs using data 
taken at centre of mass energies of $161$ and $172 \GeV$ gives for 
$\lambda \simeq 0.3$ a best resulting limit of $142 \GeV$ which is not yet 
competitive with the sensitivity achieved at HERA .
%
%%%%%%%%%%%%%
%

%  On the other hand, assuming a given value for the coupling 
%  $\lambda_{11}$, mass dependent limits on the branching 
%  ratio $\beta_{eq}$ can be derived.

Moving away from the BRW model, one can consider from a phenomenological
point of view scalar LQs undergoing both NC and CC DIS-like decays
($\beta_{eq} \times \beta_{\nu q'} \ne 0$) with $\beta_{eq}$ and 
$\beta_{\nu q'}$ treated as free parameters.
By gauge invariance this can only be relevant for LQ coupling to 
$e^+ \bar{u}$. H1 has derived~\cite{H1ICHEP533} such constraints. 
For $\beta_{\nu q'} = 90 \%$ and $\beta_{eq} = 10 \%$, LQ masses below 
$210 \GeV$ are found to be excluded at $95 \%$ CL for a coupling 
$\lambda_{11} \simeq 0.3$.
This extends far beyond the domain excluded by TeVatron 
experiments~\cite{D01GENE,CDF1GENE} which for such small values of 
$\beta_{eq}$ only exclude scalar LQ masses below $\simeq 110 \GeV$.
% --------------- FIGURE 13 : Limits beta, with D0 limit -----------------
\begin{figure}[bt]
  \begin{tabular}{p{0.35\textwidth}p{0.65\textwidth}}
  \caption[]{ \label{fig:limibeta}
  {\small Mass dependent exclusion limits at $95 \%$ CL on 
           the branching ratio $\beta_{eq} = BR(LQ \rightarrow eq)$
          for scalar leptoquarks produced by (a) $e^+ d$ and
          (b) $e^+ u$ fusion.
          Two exclusion regions (light dotted grey) corresponding to 
	  $\lambda=0.1$ and $\lambda = 0.05$ are represented.
          The $D\emptyset$ limit is also shown as hatched region.}}
 &
   \raisebox{-150pt}{\mbox{\epsfxsize=0.63\textwidth 
                 \epsffile{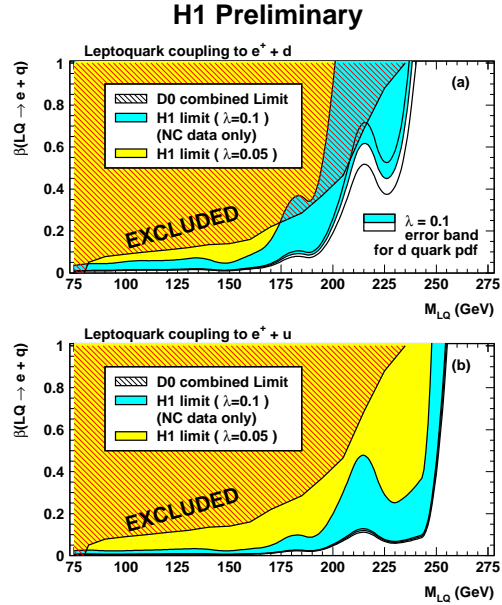}}}
 \end{tabular}
% \vspace*{-1.1cm}
\end{figure}
%-------------------------------------------------------------------------
Given a value of $\lambda_{11}$, upper limits on $\sigma_{LQ}$ 
can be translated for narrow intrinsic widths in terms of mass 
dependent limits on the branching $\beta_{eq}$ without making 
specific assumptions on the nature of the other decay modes.
Results are shown in Fig.~\ref{fig:limibeta}a and b for LQs produced 
via $e^+d$ and $e^+u$ fusion respectively.
Despite the small $\lambda_{11}$ values considered, the exclusion
domains are seen in Fig.~\ref{fig:limibeta} to extend beyond the 
region covered by the $D\emptyset$ experiment~\cite{D01GENE}
at TeVatron for small $\beta_{eq}$ even in the less favourable case of 
an LQ coupling to $e^+ d$.
HERA rules out masses below $\simeq 210 \GeV$ if $\lambda_{11}=0.1$ 
(Fig.~\ref{fig:limibeta}b) or below $\simeq 255 \GeV$ for a interaction
of electromagnetic strength (see for instance Fig.~\ref{fig:limibetatau}a).

%
%-----------------------------------------------------------------------
\subsection{Search for leptoquarks with mixed lepton flavour couplings}
\label{sec:lqlfv}

We have seen in the previous section that the mass range of interest for 
a discovery of lepton-quark resonances at HERA is severely constrained 
by TeVatron results for LQs coupling solely to first generation fermions.
Hence, H1 has investigated~\cite{H1ICHEP533} further the case of LQs 
possessing $\lambda_{11} \times \lambda_{3j} \ne 0$. Such a LQ coupling 
to both first and third generation leptons would lead to striking
lepton flavour violating (LFV) processes 
$e^+ q \rightarrow LQ \rightarrow \tau^+ q'$.

The analysis for the $LQ \rightarrow \tau + q$ decays covers the
hadronic decays of the $\tau$ as well as the decay 
$\tau^+ \rightarrow \mu^+ \nu_{\mu} \bar{\nu}_{\tau}$ but only the
former is actually used in the exclusion limits derivation.
For the latter, the analysis requires simply an isolated track 
with transverse momentum $P_T > 10 \GeV$ linked to the primary vertex 
and a visible transverse momentum flow imbalance as measured solely 
via calorimetry of $P_{T,miss}^{vis.} > 25 \GeV$.
Only four $\mu$+jet events are found to satisfy these criteria.
All four are amongst the ``outstanding'' high $P_T$ lepton events
discussed in ref.~\cite{H1MUEV}.
None of these $\mu + X$ events has a $\mu$ candidate at $\theta_{\mu}$ 
and $\phi_{\mu}$ angles corresponding to the $\tau$ angles predicted
from kinematical constraints (the $\mu$ from the $\tau$ decay should be 
strongly boosted in the $\tau$ direction). 
Hence neither $LQ \rightarrow \mu + q$ (2$^{nd}$ generation) nor
$LQ \rightarrow \tau + q \,;\, \tau \rightarrow \mu \nu \bar{\nu}$
(3$^{rd}$ generation, $\mu$ channel) candidates are found.
%
% -------------- FIGURE 12 : Limits for LQ -> tau ----------------------
\begin{figure}[tbh]
%\vspace*{-0.2cm}

 \begin{center}
  \begin{tabular}{cc}
     \mbox{\epsfxsize=0.48\textwidth 
       \epsffile{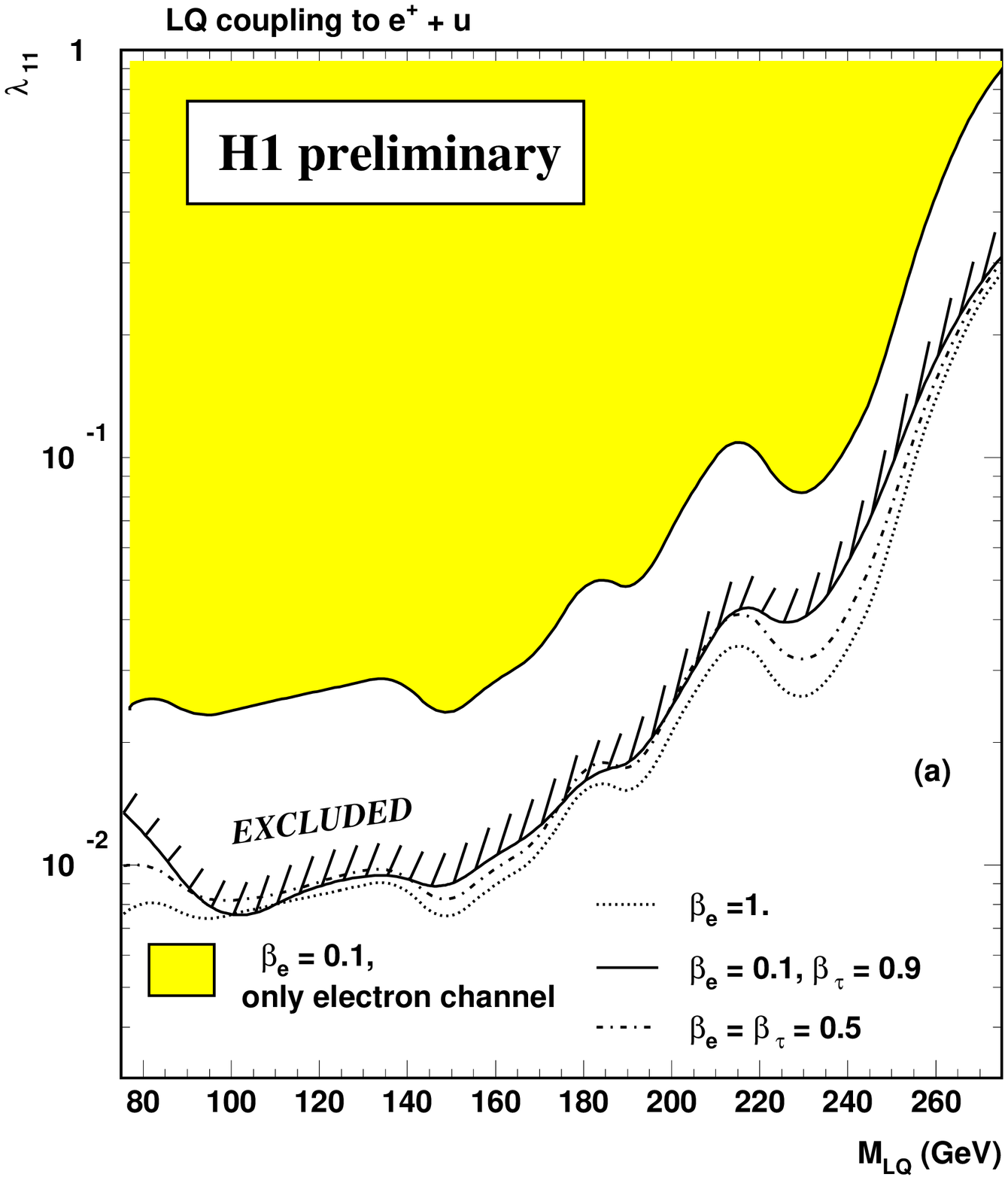}}
   &
     \mbox{\epsfxsize=0.48\textwidth 
       \epsffile{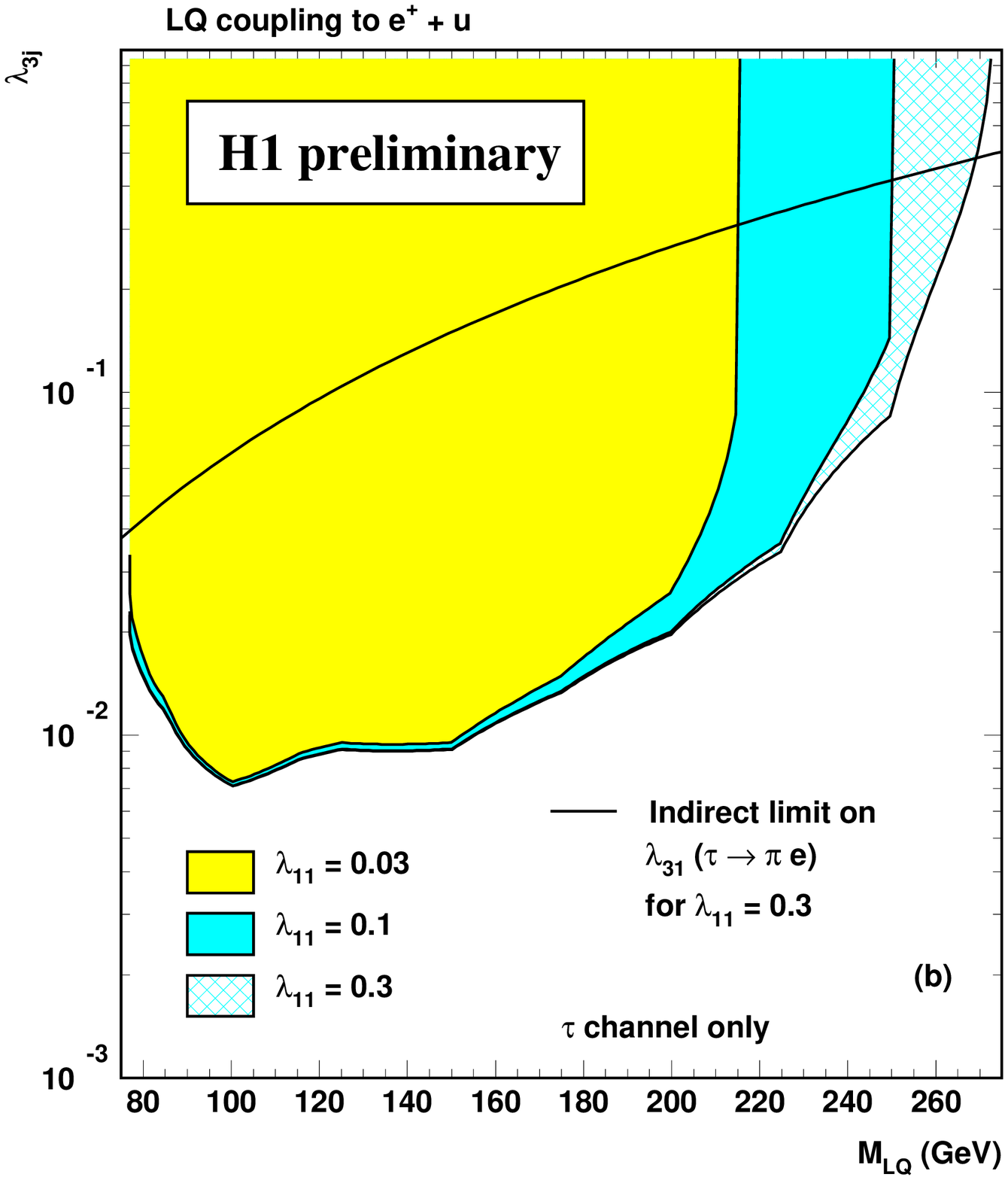}}
  \end{tabular}
 \caption[]{ \label{fig:limibetatau}
 {\small (a) $M_{LQ}$ dependent exclusion limits at $95 \%$ CL
         on the Yukawa coupling $\lambda_{11}$, for scalar
         leptoquarks produced by $e^+ u$ fusion. 
         Different hypothesis for the branching ratios into
         $eq$, $\tau q$ are considered.
         Domains above the curves are excluded. 
         (b) Exclusion domains in the plane $\lambda_{3j}$
         $(j=1,2)$ against $M_{LQ}$ for several
         fixed value of $\lambda_{11}$ (greyed areas).
         Indirect limits on $\lambda_{31}$ are represented by 
	 the dotted curve. }}
 \end{center}
\end{figure}
% ----------------------------------------------------------------------
For the hadronic decays of the $\tau$, the analysis requires a 
``pencil-like'' jet defined by a small invariant mass $M_{jet} \le 7 \GeV$
and a low multiplicity of associated charged tracks ($N_{tracks} \le 2$).
The $\tau$ jet furthermore must have either a large intrinsic $\pi^0$
component or a stiff leading track, and be back-to-back in azimuth
with the rest of the hadronic flow.
No candidate is found for these $\tau$ hadronic decay modes while 
$0.8 \pm 0.3$ misidentified background event is expected.

Mass dependent exclusion limits are shown for $\lambda_{11}$
in Fig.~\ref{fig:limibetatau}a when fixing the relative $\beta_{eq}$ and 
$\beta_{\tau q}$ branching fractions for a generic LQ coupling
to $e^+ + u$ pairs (such as the $S^\ast_{1/2,L}$ in the BRW model) 
and for three different sets of $(\beta_{eq},\beta_{\tau q})$.
Here both channels $LQ \rightarrow e^+ + jet$ and 
$LQ \rightarrow \tau^+ + jet$ are combined. This latter channel is 
essentially background free but the former benefits from a higher selection 
efficiency, such that both provide comparable sensitivity.
Hence, when $\beta_{eq} + \beta_{\tau q} \simeq 1$, the exclusion domain
is very similar to the one obtained for $\beta_{eq}=1$.
Assuming $\beta_{eq} = 10 \%$ and $\beta_{\tau q} = 90 \%$, 
masses below $275 \GeV$ are excluded at
$95 \%$ CL for $\lambda_{11} = \sqrt{4\pi\alpha}$ as shown in 
Fig.~\ref{fig:limibetatau}a.
An alternative representation of the LFV
constraints is given in Fig.~\ref{fig:limibetatau}b in the plane 
$\lambda_{3j}$ against $M_{LQ}$ for different fixed values of $\lambda_{11}$.
Here again, a LQ formed via $e^+ u$ fusion is considered 
such that only couplings $\lambda_{3j}$ with $j=1,2$ are relevant. 
Furthermore the simplifying assumption $\beta_{eq} + \beta_{\tau q} =1$ is made. 
Fig.~\ref{fig:limibetatau}b shows the domains excluded at $95 \%$ CL 
by the $\tau$ + jet final states analysis.
Of course, for each hypothetical value of $\lambda_{11}$, part of the
mass and $\lambda_{11} \times \lambda_{3j}$ domain is already implicitly
excluded anyway by the $e+q$ analysis alone.
Nevertheless, the $\tau + jet$ analysis allows extend considerably
the coverage of possible $\lambda_{11} \times \lambda_{3j}$ values;
e.g. for $\lambda_{11} = 0.03$ (1/100 of electromagnetic strength)
the mass range above $M_{LQ} \simeq 80 \GeV$ is covered in complement.
For $\lambda_{11}=\lambda_{3j}=0.03$, these new HERA results
extends the mass limits from previous HERA results~\cite{ZEUSLFV} 
by $\simeq 65 \GeV$.

The best indirect constraint~\cite{DAVIDSON} on $\lambda_{31}$, shown
in Fig.~\ref{fig:limibetatau}b in the case $\lambda_{11}=0.3$, is found
to be typically one order of magnitude less stringent than HERA 
exclusion limit.
This constraint comes from the upper limit on the branching ratio
$\beta_{\tau \rightarrow \pi^0 e}$ which could be affected by
the process $\tau \rightarrow d + LQ^*; LQ^* \rightarrow e + d$.
No low energy process constrains the coupling $\lambda_{32}$. 
Thus, a domain which was yet unexplored is now covered by H1 in the 
plane $\lambda_{32}$ against $M_{LQ}$. 
It should be noted however that other LQ species might suffer more 
severely from indirect constraints. 
This is the case in particular of LQs coupling to $e^+ d$ pairs 
(such as the $\tilde{S}_{1/2,L}$ in BRW model) for which the  
the couplings $\lambda_{31}$, $\lambda_{32}$ or $\lambda_{33}$ are 
constrained respectively by $\tau \rightarrow \pi^0 e$, 
$\tau \rightarrow K^0 e$ and $B \rightarrow \tau e X$~\cite{DAVIDSON}.
Such LQs have characteristics similar to the $\tilde{u}^j_L$
squark in SUSY models where R-parity is violated (see next
section) by couplings $\lambda'_{1j1}$ (analogous to $\lambda_{11}$) and 
$\lambda'_{3jk}$ (analogous to $\lambda_{3k}$). 
Exclusion domains which have been derived in~\cite{H1ICHEP580} in the 
plane $\lambda'_{3jk}$ against $M_{\tilde{u}}$ show that even in such
a case, HERA sensitivity extends beyond that of low energy phenomena for
masses $M_{S} \lsim 250 \GeV$.

The CDF experiment has performed a search for third generation LQ
looking at $\tau \tau b b $ final states~\cite{CDF3GENE}, and excludes a 
scalar LQ with masses below $99 \GeV$ if $\beta_{\tau b} = 1$.
A complementary search has been carried out by $D\emptyset$~\cite{D03GENE},
where the analysis of $\nu \nu b b $ final states leads to a
lower mass limit of $94 \GeV$ for $\beta_{\nu b} = 1$.
These constraints are less stringent than the $\simeq 110 \GeV$
mass limit obtained from TeVatron searches for first generation LQs
assuming $\beta_{eq} = 10 \%$.
Thus, a situation with small $\beta_{eq}$ and large $\beta_{\tau q}$ is 
especially favourable, leaving open e.g. for HERA an important discovery 
window at $M_{LQ} \lsim \sqrt{s_{ep}}$.

%=======================================================================
\section{Searches for Squarks of $R$-parity Violating Supersymmetry}
\label{sec:rpvsusy}

Squarks are scalar SUSY partners of the quarks.
%
% --- TABLE 1: Rp-violating production processes -----------------------
%
\begin{table*}[htb]
  \renewcommand{\doublerulesep}{0.4pt}
  \renewcommand{\arraystretch}{1.2}
 \begin{center}
 \begin{tabular}{p{0.40\textwidth}p{0.60\textwidth}}
         \caption
         {\small \label{tab:sqprod}
         Squark production processes at HERA ($e^+$ beam)
         via a $R$-parity violating
         $\lambda'_{1jk}$ coupling.} &
   \begin{tabular}{||c||c|c||}
   \hline \hline
   $\lambda'_{1jk}$ & \multicolumn{2}{c||}{production process} \\
   \hline
   111 & $e^+ +\bar{u} \rightarrow \tilde{d}^*_R$
       &$e^+ +d \rightarrow \tilde{u}_L $\\
   112 & $e^+ +\bar{u} \rightarrow \tilde{s}^*_R$
       &$e^+ +s \rightarrow \tilde{u}_L $\\
   113 & $e^+ +\bar{u} \rightarrow \tilde{b}^*_R$
       &$e^+ +b \rightarrow \tilde{u}_L $\\
   121 & $e^+ +\bar{c} \rightarrow \tilde{d}^*_R$
       &$e^+ +d \rightarrow \tilde{c}_L $\\
   122 & $e^+ +\bar{c} \rightarrow \tilde{s}^*_R$
       &$e^+ +s \rightarrow \tilde{c}_L $\\
   123 & $e^+ +\bar{c} \rightarrow \tilde{b}^*_R$
       &$e^+ +b \rightarrow \tilde{c}_L $\\
   131 & $e^+ +\bar{t} \rightarrow \tilde{d}^*_R$
       &$e^+ +d \rightarrow \tilde{t}_L $\\
   132 & $e^+ +\bar{t} \rightarrow \tilde{s}^*_R$
       &$e^+ +s \rightarrow \tilde{t}_L $\\
   133 & $e^+ +\bar{t} \rightarrow \tilde{b}^*_R$
       &$e^+ +b \rightarrow \tilde{t}_L $\\
   \hline \hline
  \end{tabular}
  \end{tabular}
\end{center}
\end{table*}
%.......................................................................
The most general SUSY theory which preserves gauge invariance 
of the Standard Model (SM) allows for Yukawa couplings between one 
scalar squark ($\tilde{q}$) or slepton ($\tilde{l}$) and two known 
SM fermions. Such couplings induce violation of the $R$-parity defined as 
$R_p\,=\,(-1)^{3B+L+2S}$, where $S$ denotes the spin, $B$ the baryon 
number and $L$ the lepton number of the particles.
Of special interest for HERA~\cite{DREINERH} are those Yukawa couplings
$\lambda'_{1jk}$ ($j,k$ are generation indices) violating the
leptonic number and which couple to a squark to a lepton-quark pair.
The search for $\Rp$-SUSY at HERA was carried~\cite{H1ICHEP580} 
considering otherwise the field content of the Minimal Supersymmetric 
Standard Model (MSSM) and assuming that the neutralino $\chi_i^0$ is
the lightest supersymmetric particle (LSP).

\begin{figure}[tb]
   \begin{center}
      \mbox{\epsfxsize=0.72\textwidth 
       \epsffile{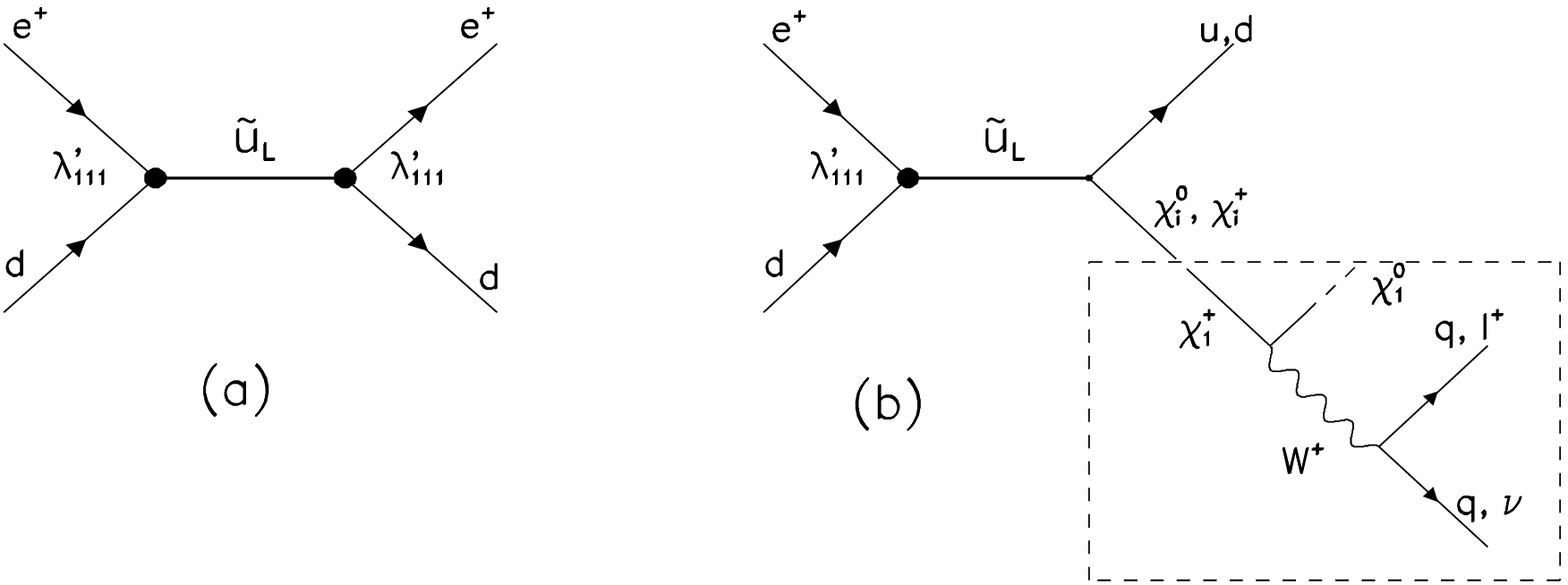}}
      \mbox{\epsfxsize=0.72\textwidth 
       \epsffile{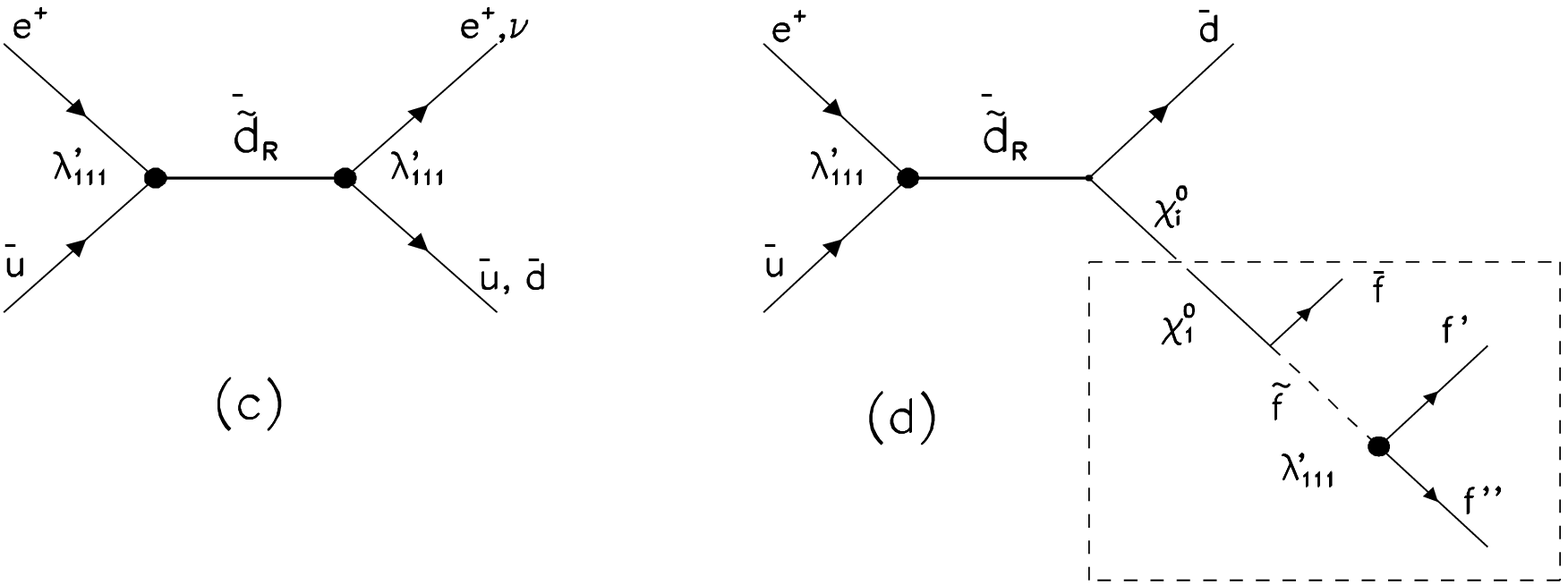}}
   \end{center}
   \caption[]{ \label{fig:sqdiag}
     {\small Lowest order $s$-channel diagrams for first generation
       squark production at HERA followed by
       (a),(c) \Rp\ decays and (b),(d) gauge decays.
       In (b) and (d), the emerging neutralino or chargino might
       subsequently undergo \Rp\ decays of which examples are
       shown in the dashed boxes for (b) the $\chi_1^{+}$ and
       (d) the $\chi_1^0$. }}
 \end{figure}

{\bf Production and decay: \,}
With an $e^+$ in the initial state, each of the nine $\lambda'_{1jk}$ 
coupling allows for a resonant production of squarks through a
$e-q$ fusion process listed in Table~\ref{tab:sqprod}. 
The squarks decay either via their $\lambda'$ coupling into SM fermions,
or via their gauge couplings into a quark and a neutralino
$\chi_i^0$ ($i=1,4$) or a chargino $\chi_j^{+}$ ($j=1,2$).
The mass eigenstates $\chi_i^0$ and $\chi_j^{+}$ are mixed states of 
gauginos and higgsinos and are in general unstable.
This latter point holds in $\Rp$-SUSY, in contrast to the strict MSSM,
also for the LSP which decays via $\lambda'_{1jk}$ into a quark, an 
antiquark and a lepton. 
In cases where both production and decay occur through a
$\lambda'_{1jk}$ coupling (e.g. Fig.~\ref{fig:sqdiag}a and c for
$\lambda'_{111} \ne 0$), the squarks behave as scalar LQs.
The ${\tilde{d}}^{k*}_R$ can decay either into $e^+ + {\bar{u}}^j$ or
$\nu_e + {\bar{d}}^j$.
Gauge invariance forbids the $\tilde{u}^j_L \rightarrow \nu q$ 
decay and, hence, such squark type is left in the \Rp\ decay mode
with $\tilde{u}^j_L \rightarrow e^+ + d^{k}$.
Hence, the final state signatures consist of a lepton and a jet and
are, event-by-event, indistinguishable from the SM neutral
and charged current DIS.
In cases where the $\tilde{u}^j_L$ (resp. ${\tilde{d}}^{k*}_R$) undergoes
a gauge decay into a $\chi^0_{\alpha}$ or a $\chi^+_{\beta}$ 
(resp. $\chi^0_{\alpha}$),
(e.g. Fig.~\ref{fig:sqdiag}b and d)
the final state will depend on the subsequent decay of the $\chi$.
Neutralinos can undergo the \Rp\ decays
$\chi^0_{\alpha} \rightarrow e^{\pm} q \bar{q}'$ or
$\chi^0_{\alpha} \rightarrow \nu q \bar{q}$, the former (latter)
being dominant if $\chi^0_{\alpha}$ is 
dominated by its photino (zino) component.
%
% --- FIGURE 13:"Feynman" Diagrams" -------------------------------------
%
%-----------------------------------------------------------------------
%
When $\chi^0_{\alpha}$ decays via \Rp\ into a charged lepton, both the
``right'' and the ``wrong'' sign lepton (compared to incident
beam) are equally probable leading to largely background free
striking signatures for lepton number violation.
On the contrary, the only \Rp\ decays for charginos are
$\chi_{\alpha}^+ \rightarrow \nu u^j \bar{d}^k$ and
$\chi_{\alpha}^+ \rightarrow e^+ d^k \bar{d}^j$. 
Neutralinos $\chi^0_{\alpha}$ with $\alpha > 1$ as well as
charginos can also undergo
gauge decays into a lighter $\chi$ and two SM fermions, through a
real or virtual gauge boson or sfermion. 

\begin{figure}[tb]
  \begin{tabular}{p{0.45\textwidth}p{0.55\textwidth}}
    \caption[]{ \label{fig:dndmS3}
    {\small Mass spectrum for $e$ + multijets final states for 
            data (symbols) and NC DIS expectation (histogram). }}
&
    \hspace*{-0.1cm} \raisebox{-150pt}{\mbox{\epsfxsize=0.55\textwidth 
                      \epsffile{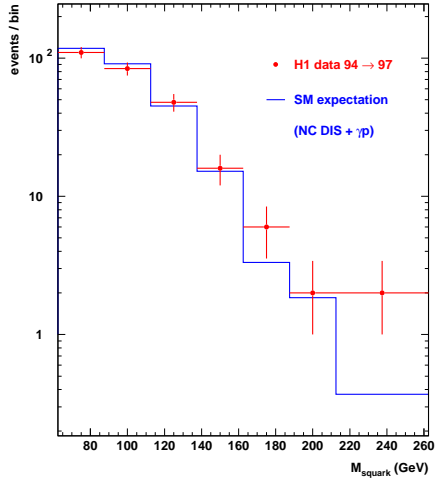}}}
 \end{tabular}
\end{figure}

{\bf Event topologies: \,}
Overall, a $e^{\pm}$+multijets final state configuration is a likely one 
for a significant part of the MSSM parameter space when considering 
squark gauge decays, as discussed in previous HERA 
analysis~\cite{HERARPV,DARKMATTER}.
Such event topologies have been searched by H1 requiring an 
isolated $e^{\pm}$ found at large $y_e$ ($y_e > 0.4$) while, nevertheless,
being accompanied by 2 forward (relative to the incident proton direction)
high $E_T$ jets.
Such a configuration is very unlikely for standard NC DIS.
H1 finds 289 candidates in the $e^{\pm}$+multijets channel, 
in good agreement with the mean SM background of $285.7 \pm 28.0$ 
expected from NC DIS and photoproduction (the latter contributes 
to less than $3 \%$).
The measured mass spectrum of the {\large{S3}} selected events is compared
to SM expectation in Fig.~\ref{fig:dndmS3}. Good agreement is observed
with only a slight excess of events observed at the highest masses.
H1 has further looked for ``wrong'' sign final state lepton in the process 
$ e^{+} q' \rightarrow \tilde{q} \rightarrow e^{-} q'' \bar{q}'' q'$.
%
% ------------------ FIGURE 14: Mass spectrum S3  --------------------------
%-------------------------------------------------------------------------
One such wrong sign candidate was observed in the data, while 
the SM prediction as estimated from Monte-Carlo was found to be 
$0.49 \pm 0.2$ (coming from NC DIS).
Hence, apart from an excess of NC-like events at large $Q^2_e$ or large
$M_e$, no significant deviation from the SM expectations has been found 
in a key topology for gauge decay channels.
Assuming that the slight deviations observed are due to statistical 
fluctuations,  the searches in \Rp\ and gauge decay channels can be 
combined to set constraints on $\Rp$-SUSY models. 

%
% ------ FIGURE : Limits on \lambda'_{1j1} vs M_squark ---------------------
%
\begin{figure}[tb]
  \begin{tabular}{p{0.35\textwidth}p{0.65\textwidth}}
 \caption[]{ \label{fig:lim_combine}
 {\small (a) Exclusion upper limits at $95\%$ CL for $\lambda'_{1j1}$ 
	as a function of the $M_{\tilde{q}}$, for a set of MSSM parameters 
	leading to a $40 \GeV$ $\chi^0_1$  dominated by its $\tilde{\gamma}$ 
	component.
        The limits are given for different possible combinations of the 
	contributing channels.
        Regions above the curves are excluded.
        (b) The relative contributions of channels $e^+$+jet ({\large{S1}}),
        $e^+$+multijets ({\large{S3}}) and $e^-$+multijets ({\large{S4}}) 
        versus $M_{\tilde{q}}$. }}  
&
    \raisebox{-190pt}{\mbox{\epsfxsize=0.65\textwidth 
             \epsffile{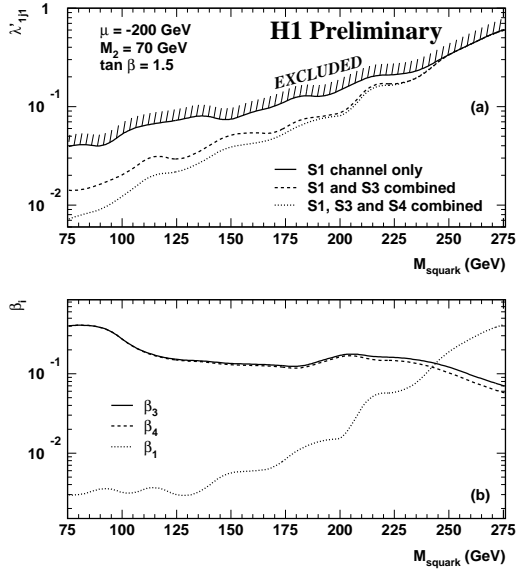}}}
  \end{tabular}
\end{figure}
%---------------------------------------------------------------------------
%
{\bf Rejection limits: \,}
The rejection limits are derived as a function of the $\tilde{u}^j_L$ 
mass assuming that only one of the $\lambda'_{1j1}$ is non vanishing
and combining all contributing channels.
The masses of other sfermions are assumed to only influence weakly the 
branching ratios of the neutralinos and charginos~\cite{DREINERH}.
Rejection limits on $\lambda'_{1j1}$ as a function of squark mass are 
shown in Fig.~\ref{fig:lim_combine}a for the $\tilde{u}^j_L$ when combining 
the relevant event topologies, taking into account either 
NC-like $e^+$+jet only, or 
NC-like $e^+$+jet combined with $e^{+}$+multijets,
or all three channels (including ``wrong sign''$e^{-}$+multijets).
The MSSM  parameters have been set here to $\mu = -200 \GeV$, 
$M_2 = 70 \GeV$ and $\tan \beta = 1.5$. With this choice of parameters, 
the lightest neutralino $\chi^0_1$ is mainly dominated by its 
photino ($\tilde{\gamma}$) component and $M_{\chi^0_1} \simeq 40 \GeV$, 
while the $\chi^+_1$ and $\chi^0_2$ are nearly
degenerate around $90 \GeV$.
Combining the three contributing channels improves the sensitivity
on $\lambda'_{1j1}$ by up to a factor $\simeq 5$ at lowest mass compared 
to the one obtained using only the NC-like channel.
The relative contributions of the three channels in the case where
$\chi^0_1$ is $\tilde{\gamma}$-like are plotted against the squark mass in 
Fig.~\ref{fig:lim_combine}b for $\lambda'$ at the current sensivity limit.
It is seen that for masses up to $\simeq 230 \GeV$, the channels 
$e^{+}$+multijets and $e^{-}$+multijets have equal and dominant 
contributions.
These channels play a decreasing role with increasing $M_{\tilde{q}}$
as squark decays into $\chi^+_1$ and $\chi^0_2$ become kinematically 
allowed. The $\chi^+_1$ and $\chi^0_2$ become dominated respectively 
by their wino and zino components~\cite{HERARPV,DARKMATTER} 
and decay preferentially into $\nu q \bar{q}$ (a channel not covered in 
ref.\cite{H1ICHEP580} except partly through CC-like analysis).
In the very high mass domain, a large Yukawa coupling is necessary
to allow squark production, hence the relative contribution of 
$e^+$+jet is largely enhanced.
Another set of values for $(\mu, M_2, \tan \beta)$ leading to a $40 \GeV$ 
$\chi^0_1$ dominated by its zino ($\tilde{Z}$) component was considered 
in~\cite{H1ICHEP580} to study the dependence of the rejection limits 
on the choice of MSSM parameters.  
In such a case the $\chi^0_1$ of the $\chi^+_1$ decay preferably in
$\nu q \bar{q}$ (rather than in $e q q'$), leading to 
multijets+$P_{T,miss}^{vis.}$ topologies not easily separable
from the SM background~\cite{HERARPV,DARKMATTER} and, hence, not expected 
to contribute very much to the sensitivity to new physics.
%.......................................................................
% --- FIGURE 7: Limits on \lambda'_{1j1} vs M_squark   -----
%.......................................................................
\begin{figure}[tb]
  \begin{tabular}{p{0.38\textwidth}p{0.62\textwidth}}
  \caption[]{ \label{fig:lim_l1j1}
  {\small  Exclusion upper limits at $95 \%$ CL for the coupling
           $\lambda'_{1j1}$ as a function of squark mass, for
           various masses and mixtures of the $\chi^0_1$;
           also represented are the most stringent indirect
           limits on $\lambda'_{111}$ and $\lambda'_{1j1}$,
           $j=2,3$. }}
 &
   \raisebox{-150pt}{\mbox{\epsfxsize=0.58\textwidth 
                     \epsffile{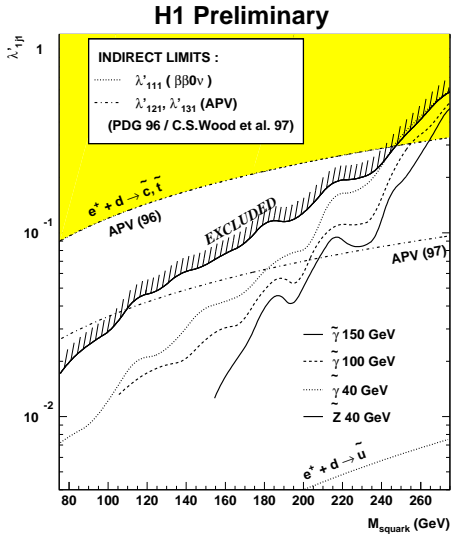}}}
 \end{tabular}
\end{figure}
%-------------------------------------------------------------------------
Since the gauge decay width of the squark does not depend
on the Yukawa coupling $\lambda'_{1j1}$, the region of the
plane $(\beta_1, M_{\tilde{q}})$ above the dotted line
in Fig.~\ref{fig:lim_combine}b is excluded at $95 \%$ CL by 
H1 combined analysis.
In particular the branching ratio of a $200 \GeV$ $\tilde{u}^j_L$
squark into $e^+ + q$ is constrained to be smaller than
$\simeq 1.5 \%$ (Fig.~\ref{fig:lim_combine}b) for the MSSM parameter
choice presented here.
It should be noted however that other specific choices of 
$(M_2, \mu, \tan \beta)$ can allow for squarks at $\sim 200 \GeV$  to 
lead to NC-like topologies with $\beta_{eq} \gsim 10 \%$. 
 
Rejection limits obtained at HERA depending on the $\chi^0_1$ mass and
nature are compared in Fig.~\ref{fig:lim_l1j1}. 
The sensitivity to $\lambda'_{1j1}$ for $M_{\tilde{q}} \lsim 200 \GeV$
is better by a factor $\simeq 2$ for a $\tilde{\gamma}$-like $\chi^0_1$
than for a $\chi^0_1$ dominated by its $\tilde{Z}$ component, due to the 
highest part of total branching actually ``seen'' in the H1
analysis~\cite{H1ICHEP580}.
One can infer from previous $\Rp$-SUSY searches at HERA~\cite{HERARPV} 
that the two cases presented here are somewhat ``extreme'' and in that 
sense quite representative of the sensitivity at HERA for any other 
choice of MSSM parameters leading to a $\simeq 40 \GeV$ $\chi^0_1$.
The sensitivity to $\lambda'_{1j1}$ for $\tilde{\gamma}$-like $\chi^0_1$
increases with $M_{\tilde{\gamma}}$ given the corresponding 
increase of efficiency for the $e$+multijets channels.
For $\lambda_{1j1} = \sqrt{ 4 \pi \alpha_{em}}$, squark masses up to 
$262 \GeV$ are excluded at $95 \%$ CL by this analysis,
and up to $175 \GeV$ for coupling strengths $\simeq 0.01 \alpha_{em}$.
For low masses, these limits represent an improvement 
of a factor $\simeq 3$ compared to H1 previously published 
results~\cite{HERARPV}. 

%
% -> Comparison with indirect limits :
%
{\bf Other indirect and direct constraints: \,}
The rejection limits obtained at HERA are compared to the best indirect 
limits in Fig.~\ref{fig:lim_l1j1}.
The most stringent indirect constraint comes from the non-observation of 
neutrinoless double beta decay~\cite{BETA0NU} but only concerns 
$\lambda'_{111}$ coupling.
The most severe indirect limits~\cite{RPVINDIR} on couplings $\lambda'_{121}$ 
and $\lambda'_{131}$, which could allow for the production of squarks 
$\tilde{c}$ and $\tilde{t}$ respectively, come from APV~\cite{APVIOL}. 
It is seen that the sensitivity at HERA is better or comparable to the most 
stringent constraints on $\lambda'_{121}$ and $\lambda'_{131}$.
For large $M_{\chi^0_1}$ values, HERA limits improves the sensitivity on
some $\lambda'$ coupling by a factor up to $\simeq 4$.

%
% -> Comparison with TeVatron results
%    --------------------------------
LQ-like searches imply stringent constraints on $\tilde{u}^j_L$ squark 
masses only if somehow $\beta_{eq}$ can be made large. 
But as explained above, this is unlikely in $\Rp$-SUSY for (say)  
$M_{\tilde{q}} \simeq 200 \GeV$.
Hence, LQ-like constraints from the Tevatron are easily evaded. 
The problem is that a small $\beta_{eq}$ is so natural in 
$\Rp$-SUSY~\cite{HIGHXYRPV} that, at first glance, only a minute portion 
of the MSSM parameter space is left if one would like to ``explain'' a 
NC-like signal at $M \simeq 200 \GeV$ in $e^+p$ collisions via the
production of a $\tilde{u}_L$-like squark (e.g. $\tilde{c}_L$ or 
$\tilde{t}_L$).
Actually, elegant solutions can be found as discussed below in the 
case of the $\tilde{t}_L$. 
At the TeVatron, SUSY searches have been mainly carried in the 
framework of minimal Supergravity (SUGRA) which imposes mass 
relations between the sparticles and $R$-parity conservation.
Recently, $D\emptyset$~\cite{D0RPV} also considered squark pair production 
leading in $\Rp$-SUGRA to like-sign dielectron events accompanied by jets, 
and has ruled out $M_{\tilde{q}} < 252 \GeV$ 
(95 \% CL) when assuming five degenerate squark flavours.
From a similar analysis by CDF~\cite{CDFRPV} restricted to 
$\lambda'_{121} \ne 0$, one can infer that a cross-section five times 
smaller would lead to a $M_{\tilde{q}}$ limit of $\simeq 150 \GeV$ depending on the 
gluino and $\chi^0$ masses. 
CDF also considered separately~\cite{CDFRPV} the pair production of a 
light stop $\tilde{t}_1$ assuming a decay into $c \chi^0_1$ and excluded 
$M_{\tilde{t}} < 130 \GeV$. 
To translate this constraint in one relevant for $\lambda'_{13k} \ne 0$, 
it should be noted that in this latter case, $\Rp$-decays of the 
$\tilde{t}$ would dominate over loop decays into $c \chi^0_1$. 
Moreover, $\Rp$-decays would themselves be negligible compared to 
$\tilde{t} \rightarrow b \chi^+_1$ decays as soon as this becomes allowed, 
i.e. if $M(\tilde{t}_1) > M(\chi^+_1)$ and if the $\tilde{t}_1$ eigenstate 
possesses a sizeable admixture of $\tilde{t}_L$. 
The subsequent decays of the $\chi^+_1$  would then lead to final states 
similar to those studied by CDF for $\tilde{t}_1 \rightarrow c \chi^0_1$.
Thus, $130 - 150 \GeV$ appears to be reasonable rough estimate of the
TeVatron sensitivity to a light $\tilde{t}$ for $\lambda'_{13k} \ne 0$. 
In summary, TeVatron and HERA sensitivities are competitive in $\Rp$-SUSY
models with five degenerate squarks, but models predicting a light
$\tilde{t}$ are better constrained at HERA provided that $\lambda'_{13j}$ 
is not too small. 

Now let's come back to the case of the $\tilde{t}_L$. 
Sizeable branching ratios for both \Rp\ and gauge decay modes for a 
$\tilde{t}$ produced via $\lambda'_{131}$ are very difficult to realize as 
argued in~\cite{ALTARELLI} when restricting to $\tilde{t}$ decays into $e+d$ 
and $b+ \chi^+_1$.
In particular, the conflicting requirements due to APV~\cite{APVIOL}
constraints (implying a lower bound for $\beta_{e q}$) and 
to direct searches at the TeVatron (implying an upper bound on 
$\beta_{e q}$) are not easily accommodated.
A special case occurs if there exists a very heavy $\chi^+_1$ 
($M_{\chi^+} > M_{\tilde{t}}$) and a light $\tilde{b}$
($M_{\tilde{b}} < M_{\tilde{t}}$) such that the $\tilde{t}$ is left with 
the decay modes $\tilde{t} \rightarrow e^+ d$ and 
$\tilde{t} \rightarrow \tilde{b} W^+$.
This interesting possibility was first discussed in~\cite{KONSBOTTOM}
as a way to ``explain'' simultaneously an excess in the NC-like channel 
and the striking observation~\cite{H1MUEV} of LFV-like events with high 
$P_{T,miss}^{vis.}$ containing a high $P_T$ muon and jet(s).
For a lightest $\tilde{b}$ mass $\simeq 100 \GeV$, simultaneous 
sizeable branching ratios $\beta_{e^+ d}$ and $\beta_{\tilde{b} W^+}$ 
become possible~\cite{SUSY98} for the $\tilde{t}$ thus extending the 
discovery potential at HERA for $M_{\tilde{t}} \simeq 200-250 \GeV$.

%=======================================================================
\section{Conclusions}
\label{sec:conclusions} 

The recent observation of possible deviations from Standard Model 
expectation in electroweak-like processes at HERA has considerably
revived the interest in new theories requiring bosons with Yukawa 
couplings to lepton-quark pairs. 
Collider experiments and low energy precision experiments have reached 
remarkable (and often comparable) sensitivity to such particles, 
providing new avenues for the manifestation of exciting new physics.

%---------------------------------------------------------------
\section*{Acknowledgments}
 
I wish to thank the organizers of the WEIN 98 Conference and the 
Chairman Prof. Cyrus Hoffman for providing me with this opportunity
to discuss the exciting prospects at HERA and other colliders
for new physics closely linked to the symmetry between the leptonic 
and quarkonic sectors of standard matter.
I wish to thank members of the H1 and ZEUS Collaborations for their 
support.
 
%---------------------------------------------------------------

%%%%%%%%%%%%%%%%%%%%%%%%%%%%%%%%%%%%%%%%%%%%%%%%%%%%%%%%%%%%%%%%%%%%%%%
%                       End of Document                               %
%%%%%%%%%%%%%%%%%%%%%%%%%%%%%%%%%%%%%%%%%%%%%%%%%%%%%%%%%%%%%%%%%%%%%%%
 
\end{document}